\documentclass[traditabstract]{aa} % for the abstract without structuration 
\usepackage{graphicx}
\usepackage{txfonts}
\usepackage{natbib}
\bibpunct {(}{)}{;}{a}{}{,} %to follow the A&A style
\usepackage{longtable}
\usepackage{multirow}
\def\km{\,km~s$^{-1}$}
\def\erg{erg cm$^{-2}$sec$^{-1}$\AA$^{-1}$}
\begin{document}
   \title{Five-Year Optical and Near Infrared Observations of the Extremely Slow Nova V1280 Scorpii\thanks{Tables 2, 3, 5 and Figures 7, 8, 10, 12
are only available in electronic form at http://www.edpsciences.org}}
   \titlerunning{Optical and Near Infrared Observations of V1280 Scorpii}
   
 \author{H. Naito\inst{1}
          \and S. Mizoguchi\inst{2}
          \and A. Arai\inst{3}
          \and A. Tajitsu\inst{4}
          \and S. Narusawa\inst{5}
          \and M. Yamanaka\inst{6}
          \and M. Fujii\inst{7}
          \and T. Iijima\inst{8}
          \and K. Kinugasa\inst{9} 
          \and M. Kurita\inst{1}
          \and T. Nagayama\inst{1}
          \and H. Yamaoka\inst{10}
         \and K. Sadakane\inst{11}        
          }
          
   \institute{Graduate School of Science, Nagoya University, Furo-cho, Chikusa-ku, Nagoya 464-8602, Japan\\
              \email{naito@stelab.nagoya-u.ac.jp}
         \and Sendai Astronomical Observatory, Nishikigaoka, Aoba-ku, Sendai 989-3123, Japan     
         \and Koyama Astronomical Observatory, Kyoto Sangyo University, Motoyama, Kamigamo, Kita-ku, Kyoto 603-8555, Japan
         \and Subaru Telescope, National Astronomical Observatory of Japan, 650 North A'ohoku Place, Hilo, HI 96720, USA
         \and Nishi-Harima Astronomical Observatory, Sayo-cho, Hyogo 679-5313, Japan
         \and Hiroshima Astrophysical Science Center, Hiroshima University, Kagamiyama, Higashi-Hiroshima, Hiroshima 739-8526, Japan
         \and Fujii Bisei Obsevatory, Kurosaki, Tamashima, Kurashiki, Okayama 713-8126, Japan
         \and Astronomical Observatory of Padova, Asiago Section, Osservatorio Astrofisico, 36012 Asiago (Vi), Italy
         \and Gunma Astronomical Observatory, Agatsuma-gun, Gunma 377-0702, Japan
         \and Graduate School of Sciences, Kyushu University, Hakozaki, Higashi-ku, Fukuoka 812-8581, Japan
         \and Astronomical Institute, Osaka Kyoiku University, Asahigaoka, Kashiwara, Osaka 582-8582, Japan
             }

\date{}

% \abstract{}{}{}{}{} 
% 5 {} token are mandatory

 \abstract
{
We present optical ($B$, $V$, $R_{\rm c}$, $I_{\rm c}$ and $y$) and near infrared ($J$, $H$ and $K_{\rm s}$) photometric and spectroscopic 
observations of a classical nova V1280 Scorpii for five years from 2007 to 2011. Our photometric observations show a  declining event 
in  optical bands shortly after the maximum light which continues $\sim$ 250 days. The event is most probably caused by a dust formation. 
The event is accompanied by a short ($\sim$ 30 days) re-brightening  episode ($\sim$ 2.5 mag in $V$),  which suggests a re-ignition of 
the surface nuclear burning. After 2008,  the $y$ band observations show a very long plateau at around $y$ = 10.5 for more than 1000 days 
until April 2011 ($\sim$ 1500 days after the maximum light). The nova had taken a very long time ($\sim$ 50 months)  before  entering 
the nebular phase (clear detection of both  [\ion{O}{iii}] 4959 and 5007) and is still continuing to generate the wind caused by H-burning. 
The finding suggests that  V1280 Sco is going through the historically slowest evolution. The interval from the maximum light 
(2007 February 16) to the beginning of the nebular phase is longer than any previously known slow novae: V723 Cas (18 months), RR Pic 
(10 months), or HR Del (8 months). It suggests that the mass of a white dwarf in the V1280 Sco system might be 0.6 $M_\mathrm{\sun}$ or smaller.
The distance, based on our measurements of the expansion velocity combined with the directly measured size of the dust shell, is estimated to 
be 1.1 $\pm$ 0.5 kpc. 
%**************************** append******************
%We estimate the mass contained 
%in the initial ejecta to be 2.3 $\pm$ 1.0 $\times$ $10^{-4}$ $M_\mathrm{\sun}$ from an analysis of [\ion{O}{i}] lines. 
%*****************************************************
} 


\maketitle

%________________________________________________________________

\section{Introduction}
It has been widely accepted that a classical nova occurs as a result of a thermonuclear runaway in a close binary system which contains 
a white dwarf (WD) and a normal star, when accreted material onto the WD bursts as it reaches the critical limit (e.g. Warner 1995, 
Bode \& Evans 2008). Novae are classified according to the decline rate and the spectroscopic features. 
In  the General Catalogue of Variable Stars (Samus et al. 2004), the parameter $t_{3}$ (number of days for a nova to decline 3 magnitudes 
from the maximum brightness) is used to classify the speed class based on the decline rate. 
When a nova is observed to have $t_{3}$ $<$ $\sim$ 100 days, it is classified as a
fast nova, 
while when a nova has $t_{3}$ $\geq$ 150 days, it is classified as a slow nova. 
Williams \cite{williams92} introduced two spectral classes; \ion{Fe}{ii} type and He/N type using early post-outburst spectra. 
\ion{Fe}{ii} type novae are characterized by prominent \ion{Fe}{ii} and other lower ionization lines with P Cygni absorption components 
and showing a slow spectral evolution. On the other hand, He/N type novae are characterized by exhibiting strong He, N and other 
high ionization lines with 
high expansion velocities and showing a very rapid spectral evolution. Slow novae tend to be \ion{Fe}{ii} type novae, 
whereas He/N novae are found only among the fast class. 

\setcounter{table}{0}
\begin{table*} %Table 1
\caption[]{List of observatories.}
\begin{tabular}{lllllc}
\hline
\hline
Photometry & & & & & \\
Observatory$\ast$		& Telescope				& Band & Instrument$\dagger$		&Observation Term		 & Nights\\
\hline
OKU				& 0.51-m reflector			& $y$, $B$, $V$, $R_{\rm c}$, $I_{\rm c}$ & 			& February 2007 -- April 2011 & 133\\
SAAO			& 1.4-m IRSF telescope		& $J$, $H$, $K{\rm s}$ & SIRPOL (1) 		& July 2009 -- August 2010	& 2\\
\hline
\hline
Spectroscopy & & & & &\\
Observatory$\ast$		& Telescope				& Resolution &Instrument$\dagger$ &Observation Term 		& Nights\\
\hline
NHAO			& 2.0-m NAYUTA telescope	& Medium	 &MALLS (2)		& February 2007 -- September 2008	& 23\\
FBO				& 0.28-m reflector			& Low 	&			& February 2007 -- February 2008	& 13\\
BAO				& 1.01-m telescope			& Medium	 &			& April 2009 			& 1\\
Subaru			& 8.2-m Subaru telescope	& High &HDS (3)			& May 2009 -- July 2011	& 9\\
GAO				& 1.5-m telescope			& Low &GLOWS		& June 2009 -- June 2010	& 5\\
HHAO			& 1.5-m KANATA telescope		& Low &HOWPol (4)		& March 2010 -- August 2010	& 3\\
KAO				& 1.3-m ARAKI telescope		& Low &LOSA/F2		& July 2010			& 1\\
Asiago			& 1.22-m telescope                     & Medium &Boller \& Chivens & April 2011                  & 1\\
\hline
\multicolumn{6}{l}{$\ast$Note:}\\
\multicolumn{6}{l}{OKU: Osaka Kyoiku University, SAAO: South African Astronomical Observatory, NHAO: Nishi-Harima Astronomical Observatory}\\
\multicolumn{6}{l}{FBO: Fujii Bisei Observatory, BAO: Bisei Astronomical Observatory, Subaru: Subaru Telescope, GAO: Gunma Astronomical Observatory}\\
\multicolumn{6}{l}{HHAO: Higashi-Hiroshima Astronomical Observatory, KAO: Koyama Astronomical Observatory, Asiago: Asiago Astrophysical Observatory}\\
\multicolumn{6}{l}{$\dagger$References:}\\
\multicolumn{6}{l}{(1) SIRPOL: Kandori et al. 2006, (2) MALLS: Ozaki \& Tokimasa 2005, (3) HDS: Noguchi et al. 2002, (4) HOWPol: Kawabata et al. 2008}\\
\end{tabular} 
\end{table*}

It had  been a problem what parameters (the WD mass, the WD magnetic field strength, the composition of the WD envelope, and the accretion rate) 
determine the nova type for a long time. 
The puzzle had been partly solved by the discovery of a universal decline law of classical novae by Hachisu \& Kato \cite{hachisu06a}. 
They found that the speed class is mainly determined by the mass of a WD, taking the optically thick wind into account (Friedjung 1966, 
Kato \& Hachisu 1994).
However, it is still difficult for any of their models to predict observed photometric behaviors near the optical peak and to describe complex spectral evolutions, 
because the model is only effective during the late (quietly) declining phase, when the flux is mainly contributed from the extended region 
emitting the free-free radiation. 
To understand the whole physical processes involved in a nova explosion, it is important to observe from the very early phase including 
the pre-maximum to the late declining phase over a long period of time. In this respect, a very slow nova provides us with a rare 
opportunity to follow its evolution closely.

V1280 Sco (Nova Scorpii 2007 No. 1) is a classical nova which was  discovered independently by Y. Nakamura and Y. Sakurai on the same night 
(2007 February 4) at 9$^{\rm th}$ visual magnitude (Yamaoka et al. 2007a). Its early spectra were taken by some observers (e.g. Naito \& 
Narusawa 2007, Kuncarayakti et al. 2008) and it was classified as an \ion{Fe}{ii} type nova (Munari et al. 2007). Yamaoka et al. \cite{yamaoka07c} 
reported that the spectrum at the pre-maximum stage looked like that of an F-type supergiant dominated by absorption lines of hydrogen, 
\ion{Fe}{ii} and other metals. It showed a somewhat slow rise to the maximum of $V$ =3.78 mag on February 16; 11.3 days after the discovery 
(Munari et al. 2007), and became a naked eye nova since V382 Vul and V1494 Aql both recorded  in 1999. Hounsell et al. \cite{hounsell10} 
published the data set of V1280 Sco observed by the Solar Mass Ejection Imager (SMEI) on board the {\it Coriolis} satellite. 
Thanks to the high time resolution, they revealed that there were three major but short episodes of brightening before February 20, 
which had not been detected by ground based observers. More notably, this nova showed a remarkable formation of dust in the very early phase 
(Das et al. 2007). After the maximum light, it faded steadily for about 12 days, followed by a precipitous decline in the visual region caused by 
a dust formation, which was directly detected by Chesneau et al. \cite{chesneau08} via VLTI near-IR and mid-IR observations. Das et al. 
\cite{das08} suggested that the dust was in clumps from near IR studies of V1280 Sco. Because of the event of a dust formation, the parameter $t_3$  can no longer be applicable as an indicator of the speed class in the case of V1280 Sco.

The distance of V1280 was estimated to be 630 $\pm$ 100 pc by Hounsell et al. \cite{hounsell10} by measuring the condensation time of dust 
grains from SMEI results, assuming that the condensation temperature of 
the dust was 1200 K and the ejection velocity was $\sim$ 600 \km. 
On the other hand, Chesneau et al. \cite{chesneau08} 
derived the distance of 1.6 $\pm$ 0.4 kpc from direct observations of the size of the expanding shell. The discrepancy in distance between these two estimations was over a factor of two; this could result from the complexity 
in the physical conditions (temperature and velocity) of the dust shell.

In this paper, we present  results of long-term multi-band photometry and low, medium and high dispersion spectroscopy spanning over 
five years from 2007 to 2011. Our observations are summarized in Sect. 2. Light curves and spectral evolutions are described in Sect. 3 
and Sect. 4, respectively. We discuss the distance, the WD mass and the ejected mass of V1280 Sco in Sect. 5 and a summary is noted in Sect. 6.

\section{Observations}
A list of observatories contributed to this study is tabulated in Table 1.
\subsection{Photometry}
We conducted $B$, $V$, $R_{\rm c}$, $I_{\rm c}$ and $y$ bands photometric observations using a CCD camera mounted on a 0.51-m
reflector at Osaka Kyoiku University (OKU) from February 2007 to April 2011. The field of view of the camera was 10\arcmin $\times$ 16\arcmin. 
We obtained differential magnitudes of the object relative to a comparison star using the aperture photometry with the APPHOT package 
in IRAF\footnote[1]{IRAF is distributed by NOAO for Research in Astronomy, Inc., under cooperative agreement with the National Science
Foundation.}. We used a nearby star, HD 152819 (B4\,IV), as a comparison star. Its magnitudes of $B = 9.951 \pm 0.049$, 
$V = 9.935 \pm 0.069$, $R_{c} = 9.878 \pm 0.058$ and $I_{c} =  9.847 \pm 0.039$ were quoted from Henden \& Munari \cite{henden07}.
%************************* insert *************************
The log of the optical photometric observations and the derived $B$, $V$, $R_{\rm c}$, $I_{\rm c}$ and $y$ magnitudes with errors are given in Table 2.
%************************* insert *************************

Our optical photometric monitoring is characterized by using an intermediate band  $y$ filter, whose wavelength coverage (530-560 nm; 
produced by Custom Scientific, Inc.\footnote[2]{See http://www.customscientific.com.}) is almost free from contamination from emission 
lines of novae arising mainly from \ion{Fe}{ii} in the early stage and [\ion{N}{ii}]  
5755, [\ion{O}{iii}] 4959 and 5007 in the late stage (nebular phase). The $y$-band observations have been promoted by M. Kato and I. Hachisu to obtain $y$ light curves representing continuum fluxes of novae for which their modeled light curves can be fitted to determine the WD mass (e.g. Hachisu et al. 2006, Hachisu \& Kato 2007, Hachisu et al. 2008). 
The $y$ magnitudes were measured by comparing with the $V$ magnitude of the comparison star ($V = y =  9.935 \pm 0.069$) following Crawford 
\cite{crawford87}. The difference between $y$ and $V$
magnitudes reflects a contamination from emission lines on the continuum observed through the wide-passband of $V$.

$J$, $H$ and $K_{\rm s}$ bands photometric observations were performed with the SIRPOL mounted on the 1.4-m IRSF telescope at South African 
Astronomical Observatory (SAAO) in 2009 and 2010.  The field of view of the SIRPOL's camera was 7\arcmin.7 $\times$ 7\arcmin.7.
We obtained near IR magnitudes by comparing to 2MASS catalogue stars in the same frame field using the APPHOT package in IRAF; 
2MASS 16\,573\,775-3\,221\,553 ($J$ = 8.291 $\pm$ 0.027, $H$ = 7.683 $\pm$ 0.071 and $K$ = 7.481 $\pm$  0.018), 
2MASS 16\,573\,563-3\,219\,050 ($J$ = 9.062 $\pm$ 0.026, $H$ = 8.192 $\pm$ 0.053 and $K$ = 7.832 $\pm$  0.029), 
2MASS 16\,574\,251-3\,218\,242 ($J$ = 8.885 $\pm$ 0.020, $H$ = 8.056 $\pm$ 0.061 and $K$ = 7.744 $\pm$  0.031)  and 
2MASS 16\,575\,182-3\,222\,448 ($J$ = 7.772 $\pm$ 0.021, $H$ = 6.870 $\pm$ 0.042 and $K$ = 6.547 $\pm$  0.029) were 
used (Cutri et al. 2003).

\subsection{Spectroscopy}
Low-dispersion spectroscopic observations from 2007 to 2008 were carried out using a spectrograph mounted on a 0.28-m telescope at 
Fujii Bisei Observatory (FBO). The spectral resolution was $R$ = $\lambda$ / $\Delta$ $\lambda$ $\cong$ 600 at H$\alpha$. 
Low-dispersion spectroscopic observations from 2009 to 2010 were carried out using 1-m class telescopes and spectrographs; 
the GLOWS (Gunma LOW resolution Spectrograph and imager) mounted on a 1.5-m telescope at Gunma Astronomical Observatory (GAO), 
the HOWPol (Hiroshima One-shot Wide-field Polarimeter) mounted on the 1.5-m KANATA telescope at Higashi-Hiroshima 
Astronomical Observatory (HHAO) of Hiroshima University and  the LOSA/F2 mounted on the 1.3-m ARAKI telescope at Koyama Astronomical Observatory (KAO) of Kyoto Sangyo University. The spectral resolutions at H$\alpha$ for GLOWS, HOWPol  and LOSA/F2 were $R$ $\cong$ 500, 400 and 600, respectively. 

Medium-dispersion spectroscopic observations were carried out using three instruments, the MALLS (Medium And Low dispersion Long slit 
Spectrograph) mounted on the 2.0-m NAYUTA telescope at Nishi-Harima Astronomical Observatory (NHAO) from 2007 to 2008, a spectrograph 
mounted on a 1.01-m telescope at Bisei Astronomical Observatory (BAO) in 2009, and the Boller \& Chivens grating spectrograph mounted 
on a 1.22-m telescope at Asiago Astrophysical Observatory in 2011. The spectral resolutions at H$\alpha$ were $R$ $\cong$ 1200 for 
these three instruments.

High-dispersion spectroscopic observations were carried out using the HDS (High Dispersion Spectrograph) mounted on the 8.2-m Subaru 
telescope from 2009 to 2011. Technical details for the HDS were described in Noguchi et al. \cite{noguchi02}. The observational mode 
was the same as used in Sadakane et al. \cite{sadakane10}, which provided the resolving power at H$\alpha$ of $R$ $\cong$ 60\,000 and covered 
the wavelength region from 4130 \AA~ to 6860 \AA.

A journal of spectroscopic observations of V1280 Sco is given in Table 3. Spectral data, except for those obtained at FBO and BAO, 
were reduced in a standard manner with the NOAO IRAF package. Spectrophotometric calibrations were made using spectra of standard stars: 
HR 1544, HR 3454, HR 4468, HR 4963, HR 5501, HR 6354, HR 7950, HR 7596, HR 8634, HR 9087 and HD 117\,880 observed on the same nights.
In order to carry out the final flux calibration on each night, we used the linearly interpolated $y$-band magnitude using two consecutive 
data obtained  before and after the date of spectroscopic observation. This is  because our raw spectroscopic data contained large 
uncertainties in absolute fluxes of both the target and standard stars due to the slit loss caused by a telescope guiding error and
varying weather and seeing conditions during spectroscopic observations. We used the translation function between the absolute flux and 
the $y$ magnitude given in Gray (1998). 
Errors in the flux were reduced to $\sim$ 10\% by using this method.
We used the value of $E(B-V)$ = 0.4 for correcting the interstellar extinction, which was obtained by dividing $A_{v}$= 1.2 (from Das et al. 2008) by the assumed total-to-selective extinction ratio $R_{v}$ = 3.1.

%which was derived by comparing a spectrum of V1280 Sco on 2007 February 14 with a spectrum of an A-type supergiant star, Deneb,  
%observed with the same instrument (see Sect. 4.2 for details). When we assumed the total-to-selective extinction ratio to be $R_{v}$ = 3.1, 
%this result agreed with $A_{v}$= 1.2 derived by Das et al. \cite{das08}.

\begin{figure*}[hbt] %Figure 1
\begin{center}
\includegraphics[scale=1.3]{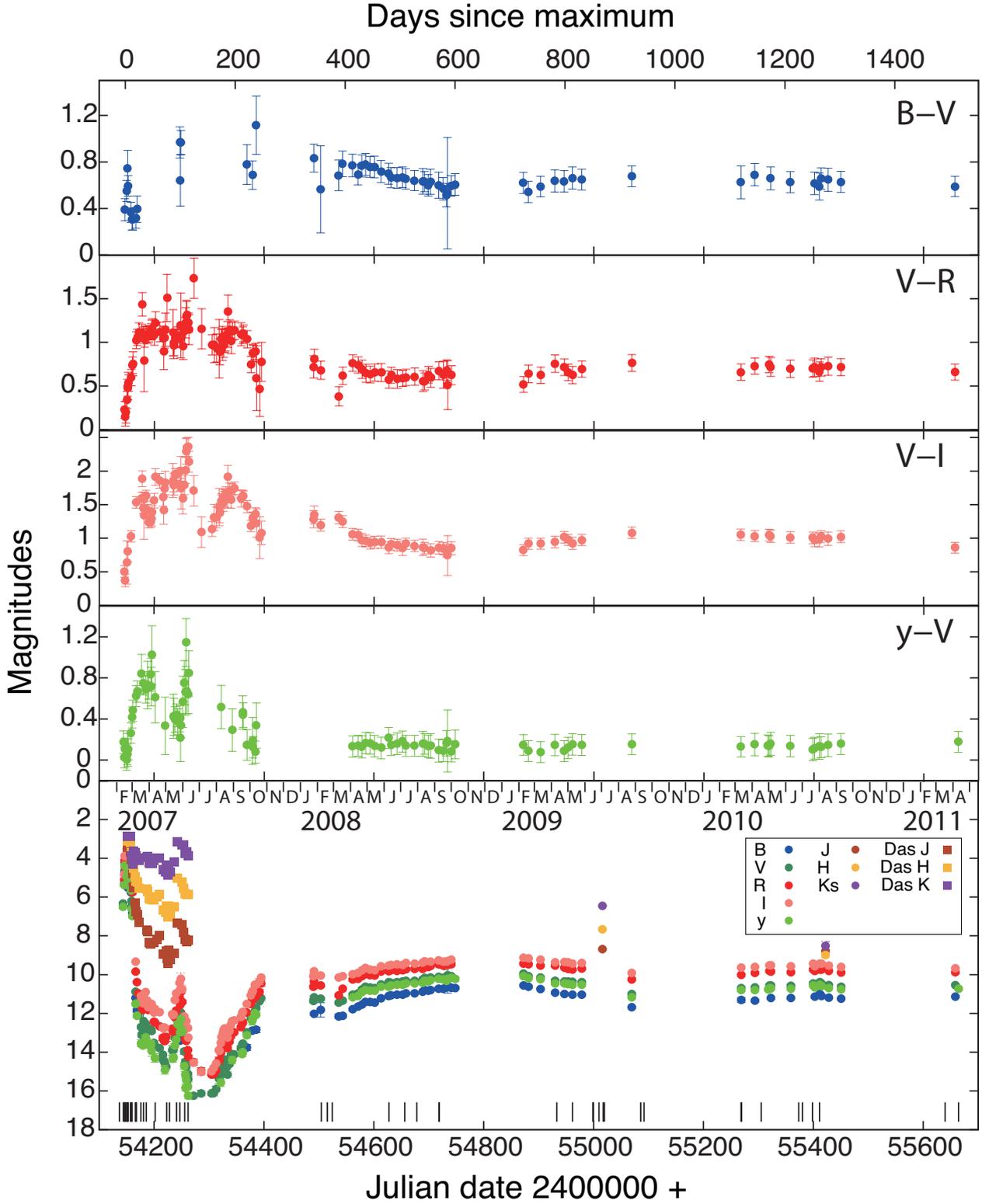}
\caption{Light curves in optical $B$, $V$, $R_{\rm c}$, $I_{\rm c}$, and $y$ and near IR $J$, $H$, and $K_{\rm s}$ bands and color indices 
spanning for five years. 
Dates of the spectroscopic observations are indicated by vertical lines.}
\label{Fig.1}
\end{center}
\end{figure*}

\section{Light Curves}
Light curves and color indices in optical wavebands and near IR light curves are shown in Fig. 1. To meet the convenience of the readers, 
near IR light curves in the early phase published by Das et al. \cite{das08} are superimposed in the figure. 
We set $t$ = 0 d on 2007 February 16 
(= JD 2\,454\,147.69), where $t$ is the time in days from the maximum brightness (Munari et al. 2007). Photometric errors in magnitudes 
are typically smaller than 0.07 mag. 

Figure 2 shows the $V$-band light curve during the initial phase (from $t$ = $-13$ d to $37$ d). We plot 
OKU $V$-band data together with the SMEI data (Hounsell et al. 2010), supplemented by several points quoted from IAU Circulars (Yamaoka et al. 2007a,b). 
The OKU $V$-band data show good agreements with the SMEI data expect for $t$ $>$ 20 d ($V$ $>$ 10), when the SMEI observations become uncertain because of its low sensitivity. 
As noted in Sect. 1, there are three episodes of short ($<$ 1 d)
brightening with amplitudes of $\sim$ 1 mag near the peak. In this paper, we call them as "peak fluctuations".

After the discovery on 2007 February 4 ($t$ = $-11.3$ d), the nova gradually increased its brightness to $V$ = 6.3 
within one week  (the initial rising phase), then remained at the pre-maximum halt stage for about two days. 
Thereafter, it reached the maximum brightness at $V_{\mathrm{max}}$ = 3.78 on 2007 February 16 ($t$ = 0 d), and underwent the peak 
fluctuations as noted above. About two weeks after the maximum, the nova entered a rapid declining phase in $V$-band until late May 2007. 

It is interesting to find that the nova remained nearly at the same brightness in $K$-band ($K$ $\sim$ 4) during the rapid decline (Fig. 3), 
and at the same time it showed sudden increases in color indices $\Delta (V-R)$ $\sim$ 1.0 and $\Delta (V-I)$ $\sim$ 1.5 as shown 
in Fig. 1. From these results, we conclude that the rapid fading spanning from early March to May 2007 had been triggered by a dust formation 
in the ejecta as reported by Chesneau et al. \cite{chesneau08} and Das et al. \cite{das08}. However, in view of the relatively small changes in the observed color indices such as $V-I$ compared to the large decline in the optical band ($\Delta V$ $\sim$ 10 within 80 days), we suspect that the observed decline can not be accounted for by the dust obscuration alone. The decline might be partly caused by the shrink of the photosphere. The clumpy nature of the dust shells as suggested by Das et al. \cite{das08} might be an important factor in accounting for the  relatively  small changes in color indices. 

In late May 2007 ($t$ $\sim$ 100 d), the first remarkable re-brightening in $V$-band occurred. 
During  the rising phase, the object showed simultaneous brightenings in $J$, $H$, and $K$ bands (Das et al. 2008). 
During the second re-brightening phase from August 2007 to October 2007, V1280 Sco was too faint to be observed through $B$ and $y$ filters 
at OKU for about two months (from $t$ = 140 d to 200 d). This implies that the $V$-band flux during the period came mainly from emission lines. 
The observation in $y$ band had recovered around $t$ = 200 d, and the index $y - V$ decreased from $\sim$ 0.5 to $\sim$ 0.2 within 50 days. 
It means that the continuum radiation revived in October 2007. At the same time, the color index $V - I$ increased from 1.1 to 1.7 
(from $t$ = 140 d to 200 d) and then decreased to 1.1 at  $t$ = 250 d. Analogous behaviors were observed in many dust-forming novae 
(e.g. V475 Sct, Chochol et al. 2005), which were interpreted that the  dust layer obscuring the continuum radiation became partly 
optically thin as dust shells expanded and spread out. Hereafter, we call the epoch from $t$ = 11 d to 250 d as the "dust phase".
 
From 2008 to 2011 (from $t$ = 420 d to 1500 d), magnitudes of $V$ and $y$ remained virtually at the same values
($y$ $\sim$ 10.5 mag) suggesting that fluxes in the optical band were contributed mainly from the photosphere
and/or free-free radiation field. On the other hand, $J$, $H$ and $K_{\mathrm{s}}$ gradually faded to
8$^{\rm th}$ magnitudes in August 2010, suggesting that the near IR radiation generated by dust particles was
low in 2010. V1280 Sco maintained its brightness until the final observation at OKU in April 2011 ($t$ = 1515 d) and showed a exceptionally long plateau spanning over 1000 days. We guess that the continuous H-burning is the primary source of the energy emitted in the optical continuum, probably because there is a sufficient amount of hydrogen gas accreted on the WD to be burned for a long time as suggested by Hachisu \& Kato \cite{hachisu06a} for slow novae.

\begin{figure}[thb] %Figure 2
\begin{center}
\includegraphics[scale=0.7]{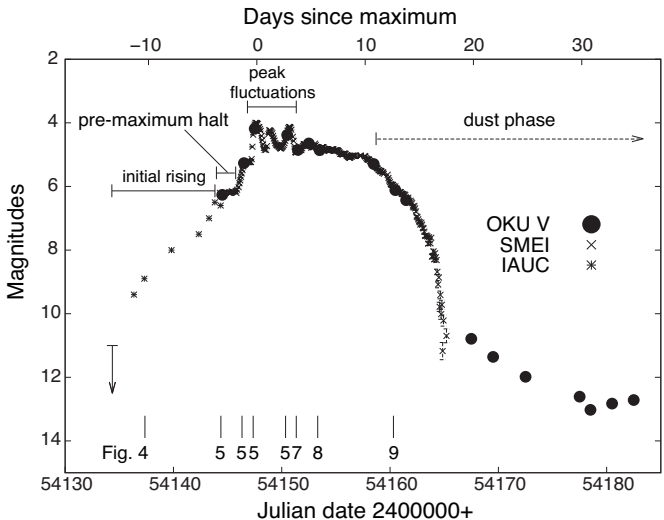}
\caption{$V$-band light curve  during the initial stage (from $t$ = $-13$ d to  37 d).
OKU data are plotted together with  the SMEI data (Hounsell et al. 2010) and those taken from 
IAU Circulars (Yamaoka et al. 2007a,b). 
There are initial rising and pre-maximum halt stages which are illustrated in a schematic light curve for nova eruptions (McLaughlin 1960). 
Peak fluctuations are observed only in the SMEI data. Dates corresponding to figures shown in this paper are indicated by vertical lines.}
\label{Fig.2}
\end{center}
\end{figure}

\begin{figure}[thb] %Figure 3
\begin{center}
\includegraphics[scale=0.7]{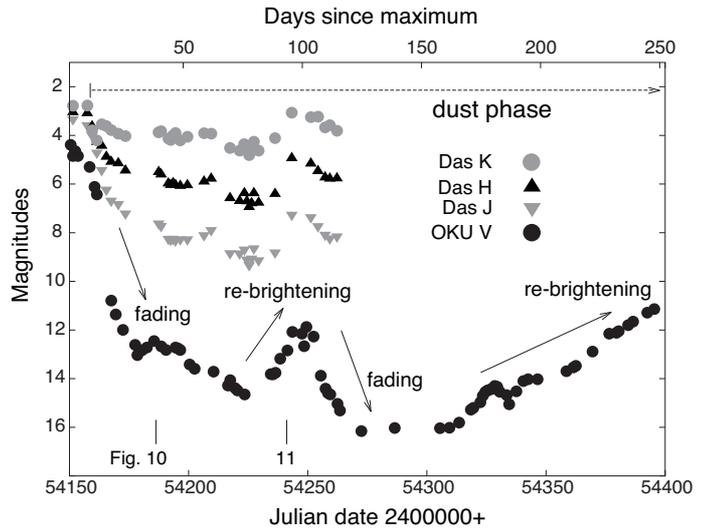}
\caption{$V$-band and $JHK$ light curves in the post-maximum phase (dust phase).   $JHK$ data points are quoted from Das et al. (2008). 
There are two episodes of fading and re-brightening. Dates corresponding to figures shown in this paper are  indicated by vertical lines.}
\label{Fig.3}
\end{center}
\end{figure}

Strope et al. \cite{strope10} defines seven classes of light curves among classical novae based on their shapes; S (smooth), 
P (plateau), D (dust dips), C (cusps), O (oscillations), F (flat-topped), and J (jitters) classes. V1280 Sco has a combination of 
D- and P-class features, although neither of these characteristics in V1280 Sco is quite prototypical when compared to  
the original classification. A typical D-class light curve shows a deep dip caused by the dust formation around 100 days after the maximum light, 
whereas the dust dip of V1280 Sco appears much earlier ($\sim$ 15-20 days after maximum light). In addition, V1280 Sco exhibits a 
re-brightening over 2-magnitudes during the dust dip period. Historically, no D-class nova had shown such a phenomenon. A typical 
P-class nova is defined to have a smooth, gradual decline in its optical light curve, then to exhibit a long-lasting plateau phase, 
which is to be followed by a steep decline to the end. According to Strope et al. \cite{strope10}, there is no nova that has a plateau 
spanning over 1000 days, with an unique exception of V1229 Aql which showed a slanted plateau with a power-law index of $-0.9$, in 
sharp contrast to that of V1280 Sco (almost flat). Furthermore, no other five classifications (S, C, O, F, or J classifications) adequately 
describes the light curve of V1280 Sco. In the light of these findings, V1280 Sco should be considered to have a very rare light curve. 
The unique light curve of V1280 Sco could provide theorists a hint to solve problems of what parameters (the WD mass, the WD magnetic
field strength, the composition of the WD envelope, and the accretion rate) determine such diverse features in light curves of novae.

\section{Spectral evolution}

\subsection{Initial rising phase}
The first spectrum taken at NHAO on 2007 February 5 ($t$ = $-10.3$ d),
 only one day after the discovery, 
is shown in Fig. 4. It displays a smooth continuum with \ion{H}{i}, \ion{Fe}{ii} (42)\footnote[3]{The multiplet number is based on Moore \cite{moore59}.}, 
\ion{Fe}{ii} (49) and \ion{Si}{ii} lines in emission, accompanied by P Cygni absorptions. Absorption lines of \ion{O}{i} and \ion{N}{i}  can also be identified. 
Observationally, V1280 Sco is a rare case for which a pre-maximum spectrum was obtained as early as 10 d before the maximum light. 
In fact, no pre-maximum spectrum had been reported in a catalog of spectroscopic observations for 27 novae (Williams et al. 2003). 
It may be a natural consequence of the very short ($<$ 5 days) rising time of classical novae (Warner 2008). 
There are several exceptional cases (V723 Cas, HR Del, V5558 Sgr, V2540 Oph, and T Pyx)\footnote[4]{See Iijima et al. \cite{iijima98} for V723 Cas; Antipova \cite{antipova78} for HR Del, Tanaka et al. \cite{tanaka11a} for V5558 Sgr, Tanaka et al. \cite{tanaka11b} for V2540 Oph and Shore et al. \cite{shore11} for T Pyx, respectively.} for which spectroscopic observations had been reported before 30 days of the maximum light or earlier. Recently, Kato \& Hachisu \cite{kato11} proposed a model to account for 
these slow novae, in which they suggested an evolution without optically thick winds. V1280 Sco might be a transitional case between 
typical classical novae which show very rapid rising time and slow novae like V723 Cas. Thanks to our spectroscopic data obtained 
on February 5, we can conclude the presence of a wind from the photosphere during the earliest phase of evolution. 

\begin{figure}[hbt] %Figure 4
\begin{center}
\includegraphics[scale=1]{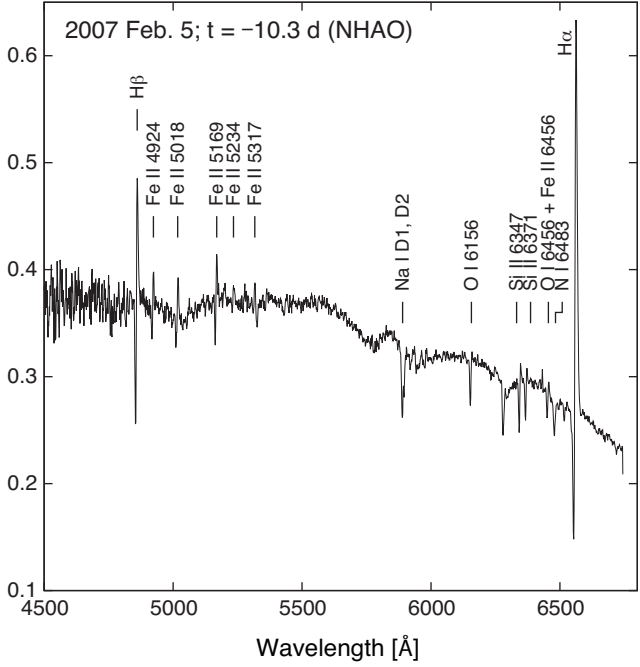}
\caption{A spectrum of V1280 Sco on 2007 February  5. Flux calibration is not applied.}
\label{Fig.4}
\end{center}
\end{figure}

\subsection{Pre-Maximum and the maximum phase}
\begin{figure}[hbt] %Figure 5
\begin{center}
\includegraphics[scale=0.7]{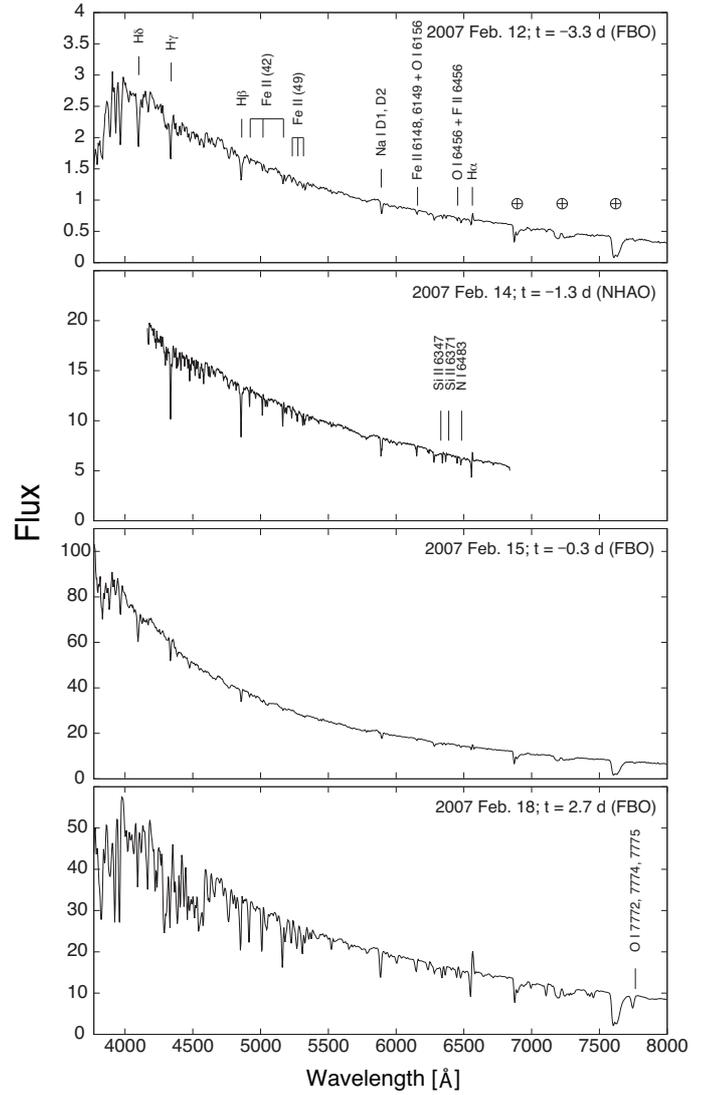}
\caption{Spectra of V1280 Sco from  2007 February 12 to 18. The unit of the ordinate is  $10^{-11}$\erg.}
\label{Fig.5}
\end{center}
\end{figure}

The flux in the optical region during the  period between 2007 February 12 and 18 (from $t$ = $-$3.3 d to 2.7 d) was mainly 
contributed by the continuum radiation coming from the photosphere and the period is called as the fireball phase (e.g. Shore 2008). 
Figure 5 shows four spectra obtained from February 12 to 18 , where the spectrum on 2007 February 12 corresponds to the pre-maximum 
halt stage and those obtained on 2007 February 15 and 18 correspond to the phase of peak fluctuations.

On 2007 February 12 ($t$ = $-$3.3 d), a very weak emission component of H$\alpha$ was detected, while all other emission lines were 
taken over by photospheric radiation to become absorption lines. On 2007 February 14 ($t$ = $-$1.3 d), the 
spectrum apparently became that of an early A-type supergiant showing weak absorption lines of \ion{He}{i} 4472 and 5876.  
On 2007 February 15 ($t$ = $-$0.3 d), near the maximum light, the effective temperature of the photosphere became the highest 
because the gradient of the continuum in the blue - visual region was the steepest. After February 15, the slope of the 
photospheric spectrum became less steep and the emission line of H$\alpha$ became strong on 2007 February 18 ($t$ = 2.7 d). 
At the same time, metallic absorption lines became much deeper than in the spectrum on 2007 February 15, especially, 
\ion{O}{i} 7772, 7774 and 7775 lines were notable. The color index $B - V$ measured on February 15 at OKU was 0.39 $\pm$ 0.09. 
When an interstellar correction of $E(B-V)$ = 0.4 was applied, the intrinsic color was $(B-V)_0$ = 0.0 $\pm$ 0.1, 
corresponding to that of an early A type star. 

Radial velocities measured from blue-shifted absorption components showed the smallest values (from $- 100$ to $- 300$ \km) on 
2007 February 14, and then the velocities became larger after the maximum light. Observed changes in radial velocities for 
various elements are displayed in Fig. 6. Similar time variations in the measured radial velocities had been found in other classical 
novae (e.g. Cassatella et al. 2004, Munari et al. 2008). There is an apparent trend that \ion{O}{i} and \ion{Si}{ii} show lower
velocities than those of Balmer and \ion{Fe}{ii} lines; mean velocities of H$\alpha$, \ion{Fe}{ii}, \ion{O}{i} and \ion{Si}{ii} 
are 620 $\pm$ 230 \km, 510 $\pm$ 150 \km, 320 $\pm$ 130 \km and 380 $\pm$ 100 \km, respectively. 
G{\l}{\c e}bocki \cite{glebocki70} found a similar trend among Balmer,  \ion{Ca}{ii} and \ion{Fe}{ii} lines in the slow 
nova HR Del. He concluded that the differences in the radial velocity came from the fact that different lines were formed at 
different heights in the expanding envelope, in which a gradient of velocity existed. For example, strong Balmer lines were formed 
in an outer layer than weaker lines of ionized metals. Consequently the velocity of expansion in this part of the envelope where 
absorption lines were formed was increasing (accelerating) with height.

\begin{figure}[hbt] %Figure 6
\begin{center}
\includegraphics[scale=0.7]{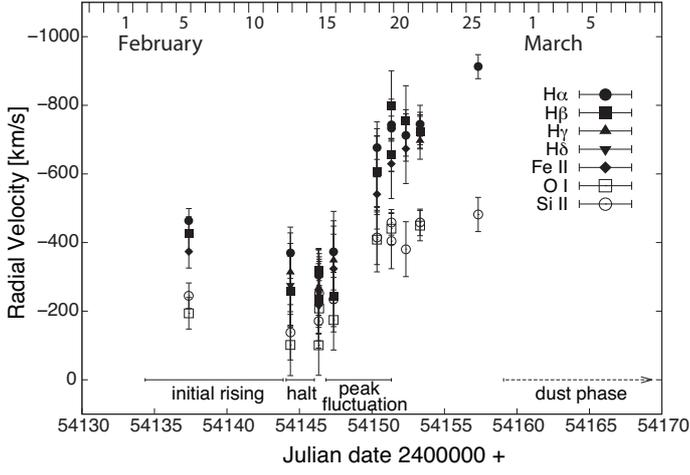}
\caption{Radial velocities of absorption components.}
\label{Fig. 6}
\end{center}
\end{figure}

\onlfig{7}{
\begin{figure*}[hbt] %Figure 7
\begin{center}
\includegraphics[scale=1]{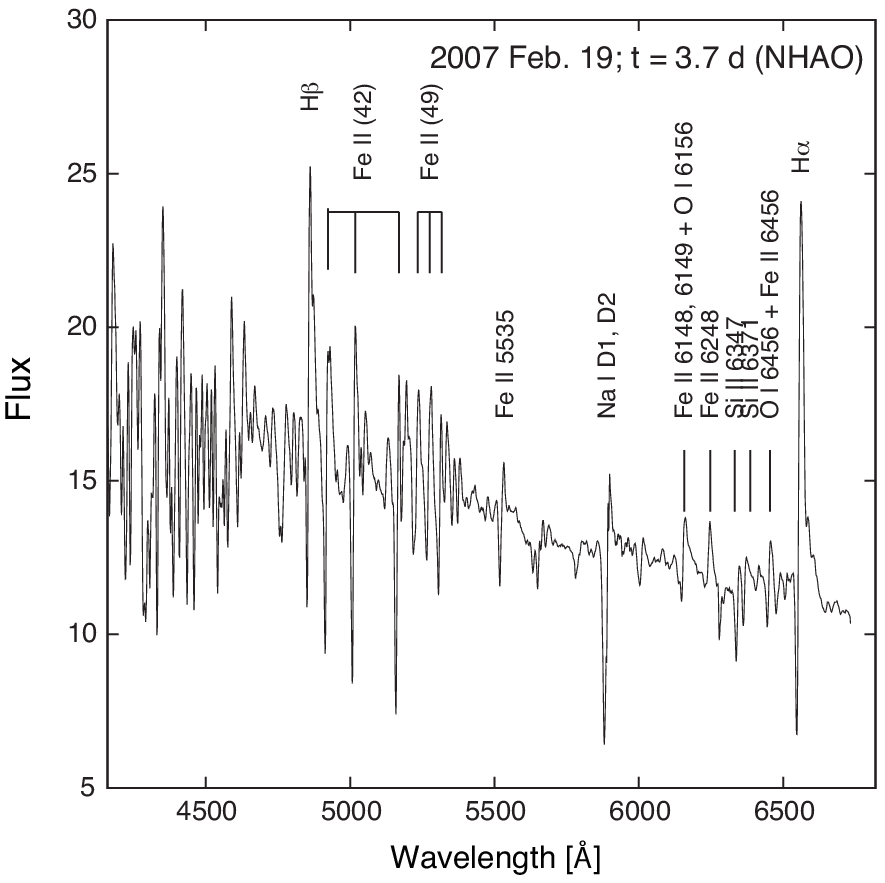}
\caption{A spectrum of V1280 Sco on 2007 February 19. The unit of the ordinate is  $10^{-11}$\erg. }
\label{Fig. 7}
\end{center}
\end{figure*}
} 

Figure 7 shows the spectrum obtained at NHAO on 2007 February 19 ($t$ = 3.7 d). It displays a typical spectral feature of  
\ion{Fe}{ii} type novae around the maximum light. The continuum spectrum weakened and emission lines of H$\beta$, \ion{O}{i}, 
\ion{Fe}{ii} and \ion{Si}{ii} became strong and they exhibited P Cygni profiles. On the night of February 19, the \ion{Na}{i} 
D1, D2 lines appeared in emission for the first time.

\onlfig{8}{
\begin{figure*}[hbt] %Figure 8 online
\begin{center}
\includegraphics[scale=1]{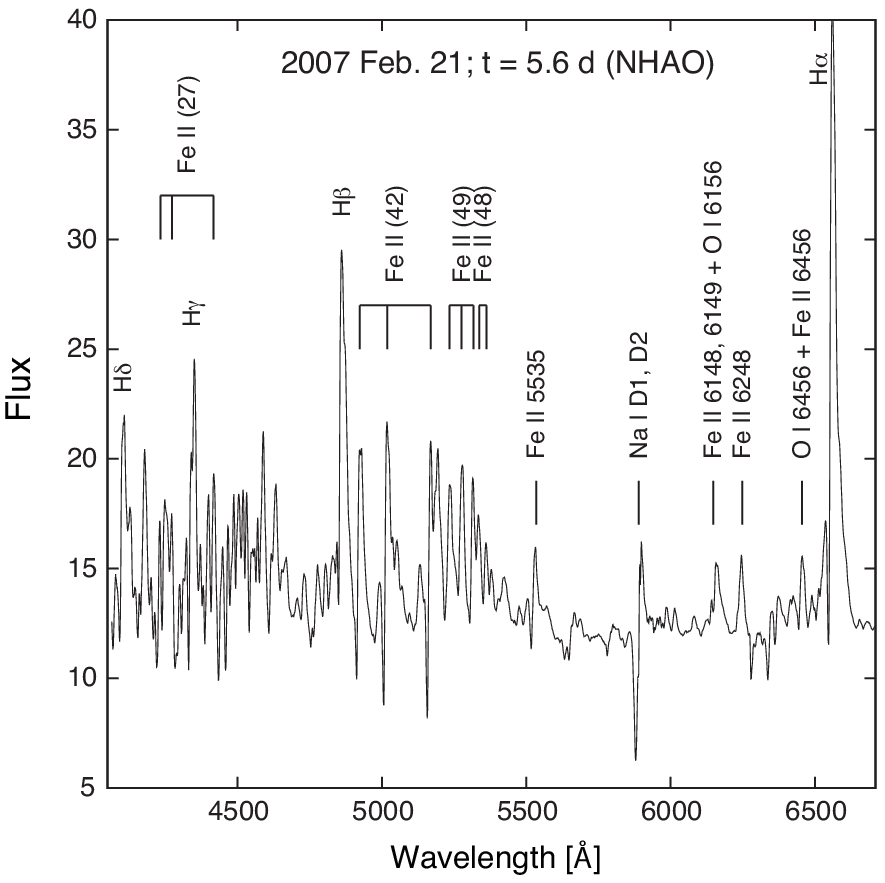}
\caption{A spectrum of V1280 Sco on 2007 February 21. The unit of the ordinate is  $10^{-11}$\erg.}
\label{Fig. 8}
\end{center}
\end{figure*}
} 

As shown in Fig. 8, the photospheric component significantly weakened and the continuum was flat on 2007 February 21 ($t$ = 5.6 d). 
This means that the free-free radiation contributed to the continuum flux dominantly as the photosphere shrunk and the 
region of optically thin gas extended at the same time. 
Absorption components became weaker, especially  the minimum point of the  H$\alpha$ 
absorption reached the local continuum level.

\subsection{Dust phase}
Forbidden lines of [\ion{O}{i}] 5577 and [\ion{O}{i}] 6300, 6364 were first observed on 2007 February 28 ($t$ = 12.6 d) 
in our observations (Fig. 9). The shape of permitted lines such as \ion{H}{i} and \ion{Fe}{ii} changed into a double-peaked 
structure. The strengths of double peaks in each Balmer line were different; the blue components were stronger than 
 the red components in H$\alpha$ and H$\beta$, while the blue components were weaker in H$\gamma$ and H$\delta$.

\begin{figure}[hbt] %Figure 9
\begin{center}
\includegraphics[scale=1]{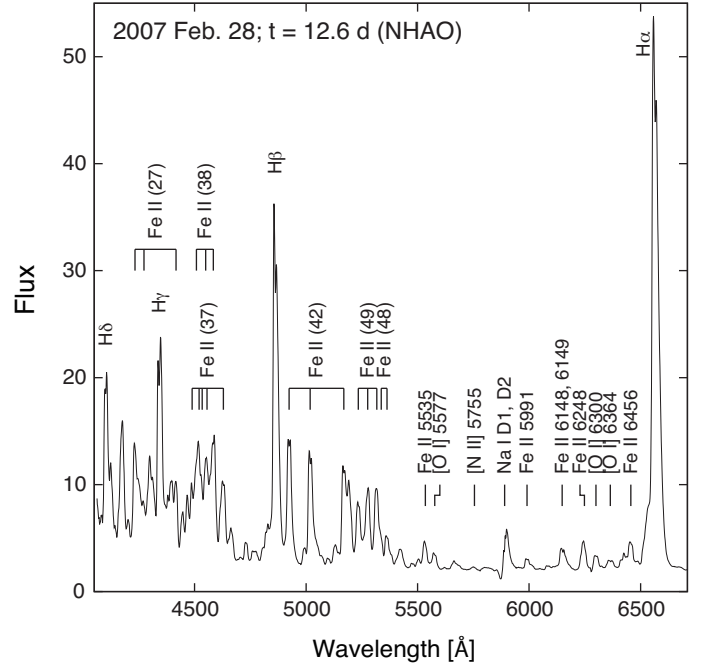}
\caption{A spectrum of V1280 Sco on 2007 February 28. The unit of the ordinate is  $10^{-11}$\erg.}
\label{Fig. 9}
\end{center}
\end{figure}

\onlfig{10}{
\begin{figure*}[hbt] %Figure 10online
\begin{center}
\includegraphics[scale=1]{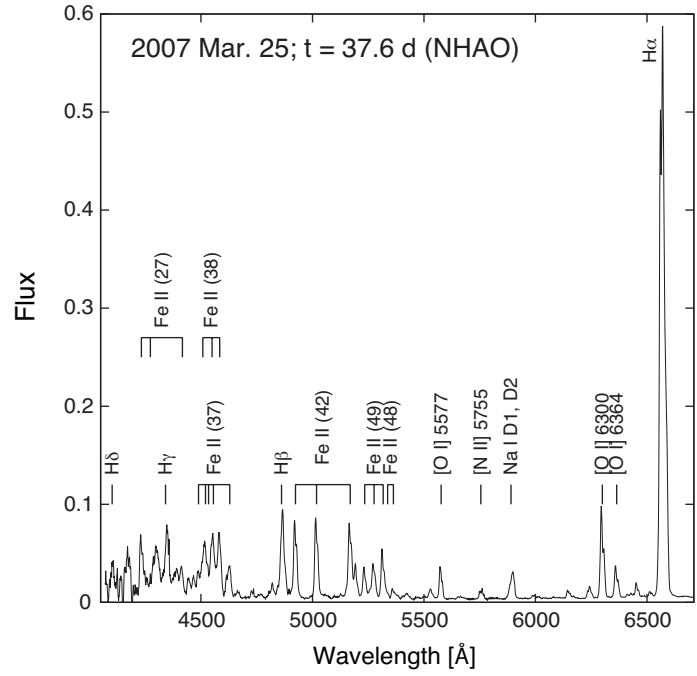}
\caption{A spectrum of V1280 Sco on 2007 March 25. The unit of the ordinate is  $10^{-11}$\erg.}
\label{Fig. 10}
\end{center}
\end{figure*}
}

\begin{figure}[hbt] %Figure 11
\begin{center}
\includegraphics[scale=1]{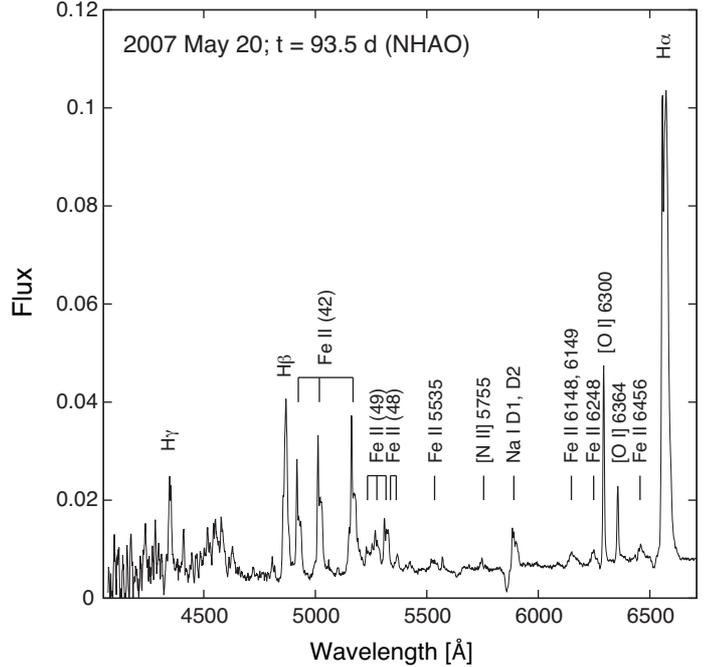}
\caption{A spectrum of V1280 Sco on 2007 May 20. The unit of the ordinate is  $10^{-11}$\erg.}
\label{Fig. 11}
\end{center}
\end{figure}

\onlfig{12}{
\begin{figure*}[hbt] %Figure 12 online
\begin{center}
\includegraphics[scale=1]{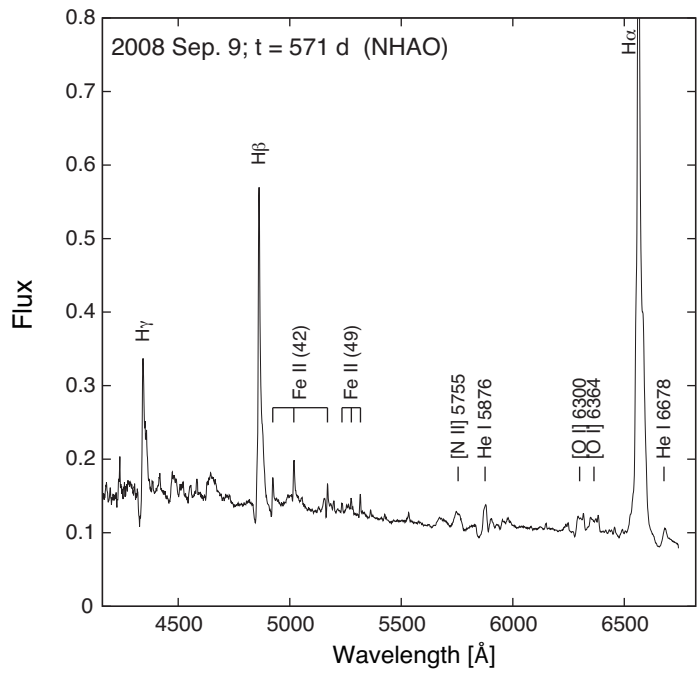}
\caption{A spectrum of V1280 Sco on 2008 September 9. The unit of the ordinate is  $10^{-11}$\erg.}
\label{Fig. 12}
\end{center}
\end{figure*}
}

\begin{figure}[hbt] %Figure 13
\begin{center}
\includegraphics[scale=0.7]{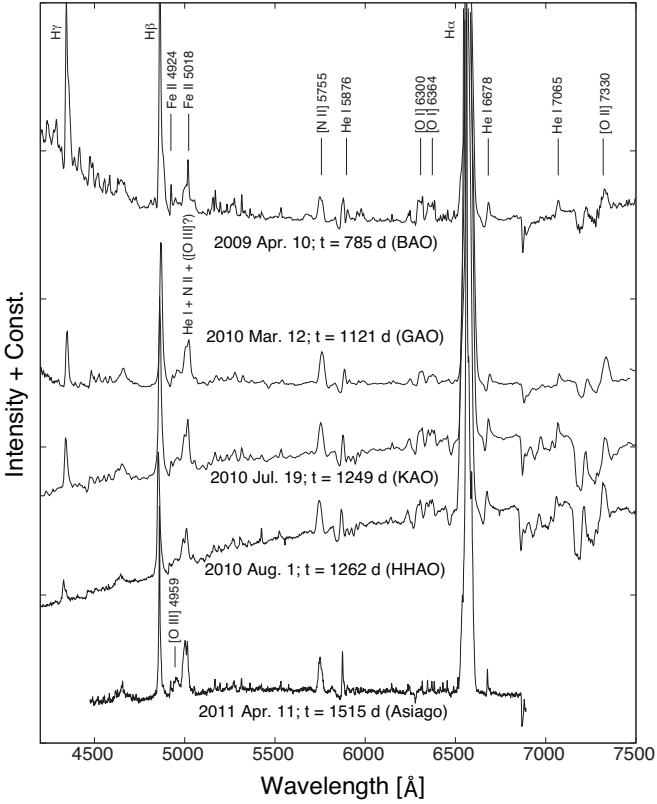}
\caption{Spectra of V1280 Sco from 2009 to 2011.}
\label{Fig. 13}
\end{center}
\end{figure}

Figure 10 shows a tracing of a spectrum obtained at NHAO on 2007 March 25 ($t$ = 37.6 d). On the spectrum, 
H$\beta$ has weakened significantly and its strength is now comparable  to those of \ion{Fe}{ii} (42) lines. Furthermore, 
the double-peaked structure of H$\alpha$ had changed dramatically between 2007 February 28 and March 25. 
Its red peak was stronger on March 25, reversing the $V$/$R$ ratio within $\sim$ 25 days.

When the decline of the brightness in the optical band began due to the dust formation from March to mid-May 2007,
the continuum flux was too weak to be detected by our instruments and the H$\beta$, H$\gamma$ and H$\delta$ emission 
lines could not be detected, too. However, during the re-brightening phase (from $t$ = 85 d to 110 d), we succeeded in 
obtaining a spectrum with high $\it S/N$ ratio in the continuum. Figure 11 shows a spectrum obtained at NHAO on 2007 May 
20 ($t$ = 93.5 d), which was the first one after the dust formation event had initiated. On the May 20 spectrum, 
H$\beta$ and H$\gamma$ lines reappeared. 
On the same spectrum, the H$\alpha$ line exhibited a double-peaked structure accompanied by a P Cygni absorption feature 
extending up to 2000 \km \, in velocity. The red peak of the double-peaked H$\alpha$ emission became much broader (FWHM $\sim$ 1200 \km). These might be connected with the second large outflow triggered by a new burst on the surface of the WD in the re-brightening period.

\subsection{Plateau from 2008 to 2010}
As shown in Fig. 12, we find that \ion{He}{i} 5876 and \ion{He}{i} 6678 appeared in emission on 2008 September 9 ($t$ = 571 d) 
for the first time. This indicates that the photospheric temperature was higher than that in 2007, however, the temperature was 
not high enough to produce [\ion{O}{iii}] or \ion{He}{ii} lines. Forbidden lines of [\ion{N}{ii}] 5755 and [\ion{O}{i}] 6300, 6364 
were broader than in 2007. P Cygni absorptions appeared in Balmer lines and \ion{Fe}{ii} lines, indicating that wind was outflowing 
due to continuous H-burning.

Figure 13 shows tracings of five spectra obtained at BAO on 2009 April 10 ($t$ = 785 d), at GAO on 2010 March 12 ($t$ = 1121 d), 
at KAO on 2010 July 19 ($t$ = 1249 d), at HHAO on 2010 August 1 ($t$ = 1262 d) and at Asiago on 2011 April 11 ($t$ = 1515 d). 
The [\ion{O}{ii}] 7330 line could be detected owing to the wider wavelength coverage at BAO, GAO, KAO, and HHAO. Spectra during 
the period were dominated by the continuum flux originated probably from the ongoing free-free radiation lasting for more than 
five years. 
There was almost no clear spectral variation between 2009 and 2011 showing appearances or disappearances of major emission lines. 
There was an emission line near 5010 \AA~ from $t$ = 785 d to 1262 d on Fig. 13. We interpret the line was mainly due to  \ion{Fe}{ii} 5018 and 
\ion{He}{i} 5016, and not due to the forbidden line of [\ion{O}{iii}]  5007.
This is because the weaker component  [\ion{O}{iii}] 4959 was absent before $t$ = 1262 d. 
However, we notice that the nebular lines of [\ion{O}{iii}] 4959 and 5007 were developing very slowly; [\ion{O}{iii}] 4959 
was detected on a spectrum taken at Asiago on 2011 April 11. The transitional period to the nebular phase will be reported in 
detail in the next sub-session. \ion{He}{I} 5876 and 6678 lines were observed with P Cygni absorptions indicating the wind was ongoing 
until April 2011.

\subsection{Onset of the nebular phase in 2011}
High resolution spectra from 4800 \AA~ to 5050 \AA~  obtained with the Subaru telescope from March to July 2011 are displayed in 
Fig. 14. This figure evidences that nebular lines of [\ion{O}{iii}] 4959 and 5007 were definitely developing during the period. 
To compare with the slow nova V723 Cas, we followed a definition of the onset of the nebular phase noted in Iijima \cite{iijima06} 
that $\it both$ [\ion{O}{iii}]  4959 and 5007 lines were identified definitely. Therefore, we conclude that V1280 Sco had entered 
the nebular phase between March and April 2011, more than 1500 d (50 months) after the maximum light. This means that it took a very long time before the photosphere to shrink so as to become hotter and emit enough UV radiation to excite doubly ionized oxygen atoms into the second energy level of 35.1 eV. However, even hotter radiation field is needed to produce \ion{He}{ii} lines for which an energy of 54.4 eV is needed. 
We detected no trace the \ion{He}{ii} 4686 line on a spectrum obtained on 2011 July 24 with the Subaru telescope. 
Considering that the historically longest transitional time to enter the nebular phase was about 18 months observed in  V723 Cas (Iijima 2006), we conclude that V1280 Sco is going through the slowest spectral evolution among known classical novae.

\begin{figure}[tbh] %Figure 14
\begin{center}
\includegraphics[scale=0.5]{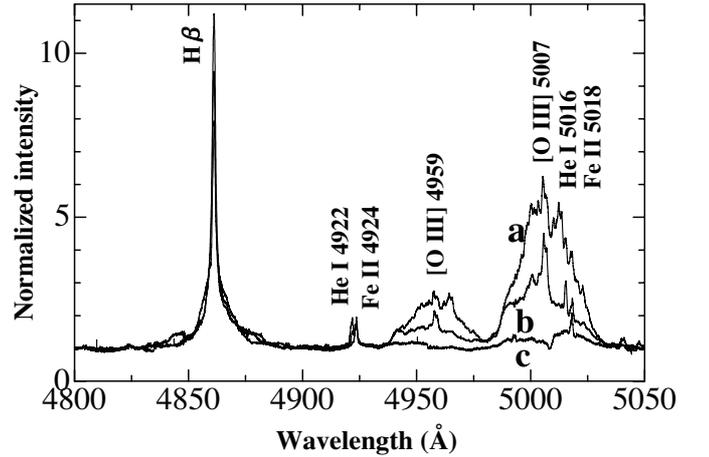}
\caption{High dispersion spectra from 4800 \AA~ to 5050 \AA~ observed with the Subaru telescope in 2011. H$\beta$ and the nebular lines  
[\ion{O}{iii}] 4959, 5007 are displayed. Labels a, b, and c represent data obtained on July 24 ($t$ = 1619 d), June 12 ($t$ = 1577 d), 
and March 17 ($t$ = 1490 d), respectively. The [\ion{O}{iii}] 4959 line is absent on March 17, while it can be clearly seen on June 12.}
\label{Fig. 14}
\end{center}
\end{figure}

\section{Discussion}
\subsection{Distance and mass of the WD}
The distance to V1280 Sco was estimated to be 1.6 $\pm$ 0.4 kpc  and 630 $\pm$ 100 pc by Chesneau et al. \cite{chesneau08} and 
Hounsell et al. \cite{hounsell10}, respectively. The substantial difference found between these results came mainly from the difficulty in measuring velocities of dust shells and photosphere. Chesneau et al. \cite{chesneau08} estimated the distance by comparing the apparent size of the 
dust shell obtained from direct observations to the calculated size, where they assumed the expansion velocity of 
500 $\pm$ 100 \km. However, it is difficult to measure the velocity of the dust shells reliably because the wind 
speed was variable before the dust phase as shown in Sect. 4.2 (Fig. 6). When we adopt the velocity of the dust shell 
as 350 $\pm$ 160 \km, which is the mean velocity measured from blue-shifted absorption lines of 
\ion{O}{i} and \ion{Si}{ii}, the distance is estimated to be $d$ = 1.1 $\pm$ 0.5 kpc by adopting the apparent 
size of the dust shell reported in Chesneau et al. \cite{chesneau08}.

Next, we attempt to evaluate the mass of the WD by using the observed pre-maximum halt shown in Fig. 2. Hachisu \& Kato \cite{hachisu04} showed 
that the brightness of a nova at the pre-maximum halt is determined by its Eddington luminosity; it is governed primarily by the mass of a WD ($M_{\rm WD}$), and is a good candidate as a standard candle. 
The absolute magnitude during the pre-maximum halt is expressed by the following relation:
\begin{eqnarray}
M_{\rm V,halt} \approx -1.53(M_{\rm WD}/M_\mathrm{\sun}) - 4.26
\end{eqnarray}
for the solar composition in WD envelopes, where $M_{\rm V,halt}$ is the absolute magnitude at the pre-maximum in $V$-band. They also showed that the absolute magnitude of the pre-maxim halt slightly depends on the composition of WD envelopes. A relation for the
case X=0.35, Y=0.33, and Z=0.32 (C+O = 0.30) is given as;
\begin{eqnarray}
M_{\rm V,halt} \approx -1.75(M_{\rm WD}/M_\mathrm{\sun}) - 4.25 .
\end{eqnarray}
These relations are valid for classical novae exhibiting a pre-maximum halt in the mass range of 0.6 $\leq$ $M_{\rm WD}/M_\mathrm{\sun}$ $\leq$ 1.3.
A comparison of 
 equations (1) and (2) shows that the variation of 
$M_{\rm V,halt}$ in the case of changing the composition from the solar abundance to enrichments of C + O = 0.30 is 
at most $\sim$ 10 \%. When we use our observational results (the magnitude at the pre-maximum halt period: $m_{\rm V,halt} = 6.3$ and 
the interstellar extinction $A_{\mathrm{v}} = 1.2$) in  equations (1) and (2) using the 
distance $d$ = 1.1 kpc, we obtain the mass of the WD as 0.6 $M_\mathrm{\sun}$ from equation (1) and even smaller one 
from  equation (2). 
To determine the WD mass more accurately, further analyses including other parameters (the abundances of He and CNO in 
WD envelopes, the WD magnetic field strength, etc.) and/or extending the range of WD mass in the theory are needed. 
In any case, we infer that the mass of a WD in  V1280 Sco is most likely to be $\sim$ 0.6 $M_\mathrm{\sun}$ or 
smaller and this is consistent with being a very slow nova as our observations show.

\subsection{Mass in the shell}
%***************************CHANGE*************************
We estimate the mass of  the nova ejecta using [\ion{O}{i}] lines at $\lambda$$\lambda$5577, 6300 and 6364 for  
two cases: (1) the solar abundance of oxygen ([O/H] = 0.0) and (2) an over-abundance of oxygen ([O/H] = +1.0). 
The latter case is motivated by the observational trend that dust forming novae have enrichments of CNO elements (Gehrz 1988, and references therein).
%**********************************************************
These [\ion{O}{i}] lines are used to estimate the physical parameters and the mass contained in the shell for classical novae, which is described first in Williams \cite{williams94}. The optical depth ($\tau$) of the  $\lambda$6300 line can be obtained from the flux ratio of $F_{\lambda6300}$ to $F_{\lambda6364}$ as follows:
\begin{eqnarray}
\frac{F_{\lambda6300}}{F_{\lambda6364}} = \frac{(1 - e^{-\tau})}{(1- e^{-\tau / 3})},
\end{eqnarray}
where the ratio would approach 3 or 1 when the optical depth approaches 0 or a moderately large value, respectively. For the observed ratio of $F_{\lambda6300}$ to $F_{\lambda6364}$ = 1.3, we find that the value of $\tau$ is 4.1 on 2007 February 28, which represents the largest optical depth before the inner part of the expanding envelope is obscured by dust.
This value is in good agreement with the result of Kucnarayakti et al. \cite{kucnarayakti08}. The electron temperature ($T_{\mathrm{e}}$) can be derived through the following formula of Williams \cite{williams94}:
\begin{eqnarray}
T_{\mathrm{e}} = \frac{11200}{ \log[43\tau/(1-e^{-\tau}) \times F_{\lambda6300}/F_{\lambda5577}]} \, \mathrm{K}.
\end{eqnarray}
With the optical depth and the electron temperature determined by using equations (3) and (4), we can estimate the mass of neutral oxygen ($M_{\mathrm{\ion{O}{i}}}$) contained in the ejecta from the following relation:
\begin{eqnarray}
M_{\mathrm{\ion{O}{i}}} = 152\,{d_{\mathrm{kpc}}}^{2} \exp \bigl(\frac{22850}{T_{\mathrm{e}} }\bigr) \times \frac{\tau}{(1-e^{-\tau})}F_{\lambda6300} \, M_{\sun},
\end{eqnarray}
where ${d_{\mathrm{kpc}}}$ is the distance to the object in kpc. The equation is slightly  modified  from the original one given in Williams \cite{williams94} to apply to our spectra which had been corrected for the interstellar extinction.

Derived physical parameters from 2007 February 28 to April 11 are listed in Table 4. The results of line flux measurement 
for [\ion{O}{i}] lines, together with other major absorption and emission lines, are listed  in Table 5. 
From 2007 Feb. 28 to Apr. 11, all oxygen atoms in the V1280 Sco ejecta can be assumed to be in the neutral state because 
no [\ion{O}{ii}] line was observed in 2007 observations, and therefore the mass derived for neutral oxygen corresponds 
to the total mass of oxygen in the ejecta ($M_{\mathrm{O}}$ = $M_{\mathrm{\ion{O}{i}}}$). 
Adopting the solar chemical composition from Asplund et al. \cite{asplund09}, we can estimate the total 
ejected mass ($M_\mathrm{ejecta}$) in the first eruption to be on the order of  $10^{-3}$ -- $10^{-4}$ $M_\mathrm{\sun}$, 
%***************************Append & CHANGE**********************
which depends on the assumed abundance of oxygen.
%Taking the suggested over-abundance of oxygen into account, $M_\mathrm{ejecta}$ ($M_\mathrm{\sun}$) $\sim$ $10^{-4}$ is 
%more likely to be real and this amount is consistent with that given in Chesneau et al. \cite{chesneau08}. 
The higher mass corresponds to the case of the solar abundance of oxygen and the lower one corresponds to the case of the over-
abundance of oxygen.
%****************************************************************
In both cases, these results strengthen our view that V1280 Sco belongs to the class of slow novae because slow
novae have a tendency of showing a more massive ejecta compared to that of ordinary novae ($10^{-5}$ -- $10^{-6}$$M_\mathrm{\sun}$).
After March 7 (dust phase), values of $M_\mathrm{ejecta}$ become
much smaller than February 28. The apparent decline seems to be due to an obscuration of the emitting 
region of the [\ion{O}{i}] lines by the thick dust shells.

\setcounter{table}{3}
\begin{table}[hbt] %Table 4
\caption[]{Physical parameters}
\begin{tabular}{lccccc}
\hline
\hline
Date  & $\tau$ & $T_\mathrm{e}$(K) & $M_\mathrm{\ion{O}{i}}$ ($M_\mathrm{\sun}$) &  \multicolumn{2}{c}{$M_\mathrm{ejecta}$ ($M_\mathrm{\sun}$)} \\
 & & & & [O/H] = 0.0& [O/H] = +1.0\\
\hline
2007 & & & & &\\
Feb. 28 & 4.10 & 5500 & 1.4$\times$$10^{-5}$ & 2.3$\times$$10^{-3}$ & 2.3$\times$$10^{-4}$ \\
Mar. 7 & 0.37 & 5400 & 5.2$\times$$10^{-7}$ & 9.1$\times$$10^{-5}$ & 9.1$\times$$10^{-6}$\\
Mar. 21  & 0.15 & 5000 & 2.0$\times$$10^{-7}$ & 3.4$\times$$10^{-5}$ & 3.4$\times$$10^{-6}$\\
Mar. 25 & 0.17 & 5200 &  2.3$\times$$10^{-7}$ & 3.9$\times$$10^{-5}$ & 3.9$\times$$10^{-6}$ \\
Apr. 11  & 0.48 & 4400 & 3.8$\times$$10^{-7}$ & 6.6$\times$$10^{-5}$ & 6.6$\times$$10^{-6}$ \\
\hline
\end{tabular} 
\end{table}

\section{Summary}
Our observations have revealed that V1280 Sco is an extremely slowly evolving nova in the historical record. The claim is based on the following evidences:
\begin{enumerate}
\item V1280 Sco is declining in brightness gradually at a very slow rate from the discovery to our last photometric observation in 2011; 
\item It has a very long plateau spanning over 1000 days in the light curve;
\item It takes a very long time ($\sim$ 50 months after the burst) to enter the nebular phase;
\item The wind is ongoing at least for about 1500 days until 2011.
\end{enumerate}

Among the known slow novae, V723 Cas had spent about 18 months before entering the nebular phase (Iijima 2006) and 
GQ Mus had the H-burning turn-off time of about 3000 days from  which Hachisu et al. \cite{hachisu08} estimated 
the wind-off time of 1000 days. V1280 Sco has spent about three times longer than V723 Cas to enter the nebular 
phase and has rewritten the record of the wind-off time of GQ Mus. According to Kato \& Hachisu \cite{kato94}, 
the time scale of nova evolution has a strong correlation with the WD mass, so that the 
observations of V1280 Sco suggest that it had bursted on a very low mass WD comparable to 
0.7 $M_\mathrm{\sun}$ of GQ Mus (Hachisu et al. 2008) or 0.59 $M_\mathrm{\sun}$ of V723 Cas (Hachisu \& Kato 2004) 
or even smaller. However, because the time scale is also slightly depends on the abundances, further analyses including the effect of the abundance of V1280 Sco and/or other parameters should be carried out by theorists.

The distance is estimated to be 1.1 $\pm$ 0.5 kpc, based on measurements of velocities of the expanding shell and direct measurements of the apparent size.
When we take the observed magnitude at the pre-maximum halt stage into account, the distance corresponds to the WD mass of $\sim$ 0.6 $M_\mathrm{\sun}$ or smaller (Hachisu \& Kato 2004). 
%**************************Append*********************************
Adopting the distance of 1.1 kpc, we estimated 
%*****************************************************************
the mass in the initial ejecta to be on the order of  $10^{-3}$ -- $10^{-4}$ $M_\mathrm{\sun}$ from the [\ion{O}{i}] line analysis.
The dust formation occurred at the very early phase, which might be connected with huge amount of ejected material compared to ordinary 
novae ($10^{-5}$ -- $10^{-6}$$M_\mathrm{\sun}$).

\begin{acknowledgements}
We are grateful to Nishi-Harima Astronomical Observatory staffs and KANATA team for their supports during observations. Cooperations of students at Osaka Kyoiku University who helped  photometric observations are gratefully acknowledged. We would like to thank  Masaki Nagamori and Yukio Ueno who helped spectroscopic observations at Bisei Astronomical Observatory. We sincerely thank Izumi Hachisu and Mariko Kato for useful comments. Thanks are also due to Kuzunori Ishibashi and Shinjirou Kouzuma for careful reading of the manuscript and suggestions. This work was supported by the Global COE Program of Nagoya University "Quest for Fundamental Principles in the Universe (QFPU)" from JSPS and MEXT of Japan. M. Y. has been supported by the JSPS Research Fellowship for Young Scientists.
\end{acknowledgements}

\Online

\setcounter{table}{1}
\begin{table*} %Table 2 online
\caption[]{$B$$V$$R_{\rm c}$$I_{\rm c}$$y$ magnitudes of V1280 Sco observed at OKU.}
\begin{tabular}{lllllllllll}
\hline
\hline
JD & \multicolumn{9}{c}{Magnitudes} \\
2400000+ &  \multicolumn{2}{c}{$B$ (err.)} &  \multicolumn{2}{c}{$V$ (err.)}  &  \multicolumn{2}{c}{$R_{\rm c}$ (err.)}   & \multicolumn{2}{c}{$I_{\rm c}$ (err.)} &\multicolumn{2}{c}{$y$ (err.)} \\
\hline
54144	&		&		&	6.26 	&	0.08 	&		&		&		&		&	6.44 	&	0.08 	\\
54146	&		&		&	5.27 	&	0.07 	&	5.04 	&	0.06 	&	4.76 	&	0.05 	&	5.30 	&	0.07 	\\
54147	&	4.57 	&	0.05 	&	4.18 	&	0.08 	&	4.04 	&	0.07 	&	3.81 	&	0.05 	&	4.29 	&	0.07 	\\
54148	&		&		&	4.58 	&	0.09 	&	4.38 	&	0.09 	&		&		&		&		\\
54150	&	4.94 	&	0.05 	&	4.39 	&	0.07 	&	4.05 	&	0.07 	&	3.75 	&	0.31 	&	4.39 	&	0.07 	\\
54151	&		&		&	4.85 	&	0.07 	&	4.37 	&	0.07 	&		&		&	4.91 	&	0.07 	\\
54152	&	5.39 	&	0.06 	&	4.65 	&	0.15 	&	4.15 	&	0.06 	&	3.84 	&	0.06 	&	4.75 	&	0.07 	\\
54153	&	5.45 	&	0.05 	&	4.85 	&	0.07 	&	4.32 	&	0.06 	&		&		&	4.96 	&	0.07 	\\
54158	&	5.66 	&	0.05 	&	5.30 	&	0.07 	&	4.70 	&	0.06 	&	4.27 	&	0.05 	&	5.56 	&	0.07 	\\
54160	&	6.42 	&	0.06 	&	6.12 	&	0.07 	&	5.38 	&	0.06 	&		&		&	6.53 	&	0.07 	\\
54161	&	6.74 	&	0.05 	&	6.43 	&	0.07 	&	5.67 	&	0.12 	&		&		&	6.91 	&	0.07 	\\
54167	&	11.12 	&	0.05 	&	10.80 	&	0.07 	&	9.77 	&	0.06 	&	9.26 	&	0.04 	&	11.42 	&	0.07 	\\
54169	&	11.76 	&	0.09 	&	11.36 	&	0.07 	&	10.30 	&	0.06 	&		&		&	12.03 	&	0.08 	\\
54172	&		&		&	11.99 	&	0.08 	&	10.88 	&	0.08 	&		&		&		&		\\
54177	&		&		&	12.61 	&	0.08 	&	11.50 	&	0.06 	&	11.02 	&	0.04 	&	13.45 	&	0.18 	\\
54178	&		&		&	13.03 	&	0.11 	&	11.59 	&	0.08 	&	11.14 	&	0.04 	&		&		\\
54180	&		&		&	12.83 	&	0.07 	&	11.75 	&	0.06 	&	11.38 	&	0.04 	&	13.58 	&	0.09 	\\
54181	&		&		&	12.30 	&	0.35 	&	11.51 	&	0.08 	&	10.97 	&	0.04 	&		&		\\
54182	&		&		&	12.72 	&	0.09 	&	11.67 	&	0.06 	&	11.13 	&	0.04 	&		&		\\
54185	&		&		&	12.45 	&	0.07 	&	11.36 	&	0.06 	&	10.82 	&	0.04 	&	13.14 	&	0.11 	\\
54188	&		&		&	12.68 	&	0.07 	&	11.60 	&	0.06 	&	11.25 	&	0.04 	&	13.41 	&	0.22 	\\
54190	&		&		&	12.82 	&	0.07 	&	11.72 	&	0.06 	&	11.58 	&	0.04 	&	13.53 	&	0.09 	\\
54194	&		&		&	12.73 	&	0.07 	&	11.60 	&	0.06 	&	11.48 	&	0.04 	&	13.56 	&	0.14 	\\
54195	&		&		&	12.76 	&	0.09 	&	11.61 	&	0.06 	&	11.47 	&	0.05 	&	13.47 	&	0.16 	\\
54196	&		&		&	12.82 	&	0.09 	&	11.66 	&	0.07 	&	11.43 	&	0.05 	&	13.85 	&	0.27 	\\
54200	&		&		&	13.43 	&	0.10 	&	12.34 	&	0.06 	&	11.87 	&	0.05 	&		&		\\
54202	&		&		&	13.59 	&	0.12 	&	12.37 	&	0.06 	&	11.68 	&	0.05 	&	14.21 	&	0.22 	\\
54210	&		&		&	13.71 	&	0.09 	&	12.59 	&	0.06 	&	11.85 	&	0.06 	&		&		\\
54216	&		&		&	14.28 	&	0.13 	&	13.24 	&	0.07 	&	12.67 	&	0.05 	&		&		\\
54217	&		&		&	14.06 	&	0.17 	&	13.16 	&	0.12 	&	12.64 	&	0.12 	&		&		\\
54219	&		&		&	14.40 	&	0.11 	&	13.34 	&	0.06 	&	12.65 	&	0.04 	&		&		\\
54220	&		&		&	14.48 	&	0.16 	&	13.34 	&	0.10 	&	12.66 	&	0.06 	&	14.82 	&	0.23 	\\
54223	&		&		&	14.65 	&	0.17 	&	13.14 	&	0.21 	&		&		&		&		\\
54234	&		&		&	13.82 	&	0.24 	&	12.70 	&	0.11 	&	11.95 	&	0.07 	&		&		\\
54235	&		&		&	13.83 	&	0.13 	&	12.87 	&	0.06 	&	12.00 	&	0.05 	&	14.26 	&	0.17 	\\
54236	&		&		&	13.76 	&	0.13 	&	12.76 	&	0.07 	&	11.97 	&	0.05 	&	14.18 	&	0.12 	\\
54238	&		&		&	13.18 	&	0.08 	&	12.13 	&	0.06 	&	11.23 	&	0.04 	&	13.58 	&	0.10 	\\
54241	&		&		&	12.84 	&	0.07 	&	11.76 	&	0.06 	&	10.89 	&	0.04 	&	13.28 	&	0.08 	\\
54243	&		&		&	12.08 	&	0.07 	&		&		&		&		&	12.46 	&	0.10 	\\
54247	&	13.12 	&	0.11 	&	12.15 	&	0.08 	&	11.11 	&	0.06 	&	10.35 	&	0.05 	&	12.56 	&	0.10 	\\
54248	&	13.31 	&	0.06 	&	12.67 	&	0.21 	&	11.48 	&	0.31 	&	10.67 	&	0.05 	&	12.89 	&	0.08 	\\
54249	&	12.84 	&	0.07 	&	11.87 	&	0.08 	&	10.82 	&	0.07 	&	10.13 	&	0.30 	&	12.21 	&	0.09 	\\
54252	&		&		&	12.27 	&	0.14 	&	11.31 	&	0.07 	&	10.68 	&	0.19 	&	12.84 	&	0.21 	\\
54255	&		&		&	13.88 	&	0.14 	&	12.75 	&	0.07 	&	12.08 	&	0.04 	&	14.64 	&	0.17 	\\
54257	&		&		&	14.41 	&	0.15 	&	13.19 	&	0.09 	&	12.40 	&	0.13 	&	15.08 	&	0.17 	\\
54258	&		&		&	14.60 	&	0.17 	&	13.30 	&	0.07 	&	12.31 	&	0.09 	&	15.75 	&	0.17 	\\
54259	&		&		&	14.66 	&	0.14 	&	13.34 	&	0.07 	&	12.30 	&	0.05 	&	15.33 	&	0.17 	\\
54262	&		&		&	15.03 	&	0.09 	&	13.80 	&	0.06 	&	12.67 	&	0.11 	&	15.67 	&	0.17 	\\
54263	&		&		&	15.31 	&	0.15 	&	14.16 	&	0.09 	&	13.17 	&	0.05 	&	16.16 	&	0.17 	\\
54272	&		&		&	16.15 	&	0.17 	&	14.42 	&	0.16 	&	14.45 	&	0.16 	&		&		\\
54286	&		&		&	16.03 	&	0.17 	&	14.88 	&	0.16 	&	14.94 	&	0.16 	&		&		\\
54305	&		&		&	16.04 	&	0.10 	&	15.06 	&	0.06 	&	14.90 	&	0.05 	&		&		\\
54309	&		&		&	16.02 	&	0.18 	&	15.05 	&	0.09 	&	14.71 	&	0.08 	&		&		\\
54313	&		&		&	15.80 	&	0.14 	&	14.87 	&	0.08 	&	14.49 	&	0.07 	&		&		\\
54318	&		&		&	15.28 	&	0.27 	&	14.36 	&	0.20 	&	13.82 	&	0.06 	&		&		\\
54319	&		&		&	15.21 	&	0.12 	&	14.31 	&	0.08 	&	13.83 	&	0.08 	&		&		\\
54322	&		&		&	14.97 	&	0.09 	&	13.93 	&	0.07 	&	13.42 	&	0.04 	&	15.49 	&	0.19 	\\
54323	&		&		&	14.71 	&	0.08 	&	13.66 	&	0.08 	&	13.14 	&	0.05 	&		&		\\
54324	&		&		&	14.55 	&	0.16 	&	13.56 	&	0.12 	&		&		&		&		\\
54325	&		&		&	14.49 	&	0.08 	&	13.50 	&	0.06 	&		&		&		&		\\
54326	&		&		&	14.45 	&	0.11 	&	13.49 	&	0.07 	&	12.93 	&	0.04 	&		&		\\
54327	&		&		&	14.41 	&	0.09 	&	13.42 	&	0.06 	&	12.78 	&	0.04 	&		&		\\
54328	&		&		&	14.30 	&	0.11 	&	13.28 	&	0.07 	&	12.71 	&	0.04 	&		&		\\
54329	&		&		&	14.33 	&	0.09 	&	13.32 	&	0.07 	&	12.70 	&	0.05 	&		&		\\
54330	&		&		&	14.55 	&	0.08 	&	13.49 	&	0.06 	&	12.87 	&	0.05 	&		&		\\
54333	&		&		&	14.68 	&	0.13 	&	13.58 	&	0.07 	&	13.09 	&	0.05 	&		&		\\
54334	&		&		&	15.06 	&	0.17 	&	13.71 	&	0.09 	&	13.15 	&	0.06 	&		&		\\
\hline
 \end{tabular} 
\end{table*}

\setcounter{table}{1}
\begin{table*} %Table 2 online
\caption[]{continued.}
\begin{tabular}{lllllllllll}
\hline
\hline
JD & \multicolumn{9}{c}{Magnitudes} \\
2400000+ &  \multicolumn{2}{c}{$B$ (err.)} &  \multicolumn{2}{c}{$V$ (err.)}  &  \multicolumn{2}{c}{$R_{\rm c}$ (err.)}   & \multicolumn{2}{c}{$I_{\rm c}$ (err.)} &\multicolumn{2}{c}{$y$ (err.)} \\
\hline
54337	&		&		&	14.52 	&	0.10 	&	13.38 	&	0.06 	&	12.81 	&	0.06 	&		&		\\
54340	&		&		&	14.10 	&	0.11 	&	13.08 	&	0.09 	&	12.52 	&	0.07 	&		&		\\
54342	&		&		&	14.02 	&	0.08 	&	12.90 	&	0.06 	&	12.30 	&	0.04 	&	14.32 	&	0.19 	\\
54346	&		&		&	14.03 	&	0.08 	&	12.89 	&	0.06 	&	12.28 	&	0.04 	&		&		\\
54358	&		&		&	13.70 	&	0.11 	&	12.61 	&	0.07 	&	12.11 	&	0.06 	&		&		\\
54361	&		&		&	13.54 	&	0.08 	&	12.46 	&	0.06 	&	11.91 	&	0.04 	&	14.00 	&	0.14 	\\
54362	&		&		&	13.47 	&	0.08 	&	12.36 	&	0.06 	&	11.85 	&	0.04 	&	13.91 	&	0.09 	\\
54369	&	13.67 	&	0.15 	&	12.90 	&	0.08 	&	11.86 	&	0.06 	&	11.42 	&	0.05 	&	13.05 	&	0.13 	\\
54376	&	12.80 	&	0.10 	&	12.16 	&	0.07 	&	11.41 	&	0.06 	&	10.98 	&	0.04 	&		&		\\
54379	&		&		&	12.11 	&	0.08 	&	11.23 	&	0.06 	&	10.81 	&	0.04 	&	12.30 	&	0.22 	\\
54380	&		&		&	12.05 	&	0.07 	&	11.18 	&	0.07 	&	10.76 	&	0.04 	&	12.18 	&	0.08 	\\
54384	&		&		&	11.81 	&	0.07 	&	10.91 	&	0.06 	&	10.44 	&	0.06 	&	11.89 	&	0.08 	\\
54386	&	12.77 	&	0.14 	&	11.65 	&	0.21 	&	11.06 	&	0.31 	&	10.42 	&	0.04 	&	11.99 	&	0.08 	\\
54392	&		&		&	11.29 	&	0.27 	&	10.82 	&	0.15 	&	10.28 	&	0.15 	&		&		\\
54395	&		&		&	11.14 	&	0.16 	&	10.37 	&	0.16 	&	10.06 	&	0.07 	&		&		\\
54490	&		&		&	11.27 	&	0.10 	&	10.55 	&	0.08 	&	9.99 	&	0.09 	&		&		\\
54491	&	11.93 	&	0.07 	&	11.09 	&	0.10 	&	10.28 	&	0.06 	&	9.74 	&	0.10 	&		&		\\
54503	&	11.73 	&	0.37 	&	11.17 	&	0.08 	&	10.49 	&	0.07 	&	9.97 	&	0.04 	&		&		\\
54536	&	12.06 	&	0.11 	&	11.37 	&	0.07 	&	10.99 	&	0.08 	&	10.07 	&	0.06 	&		&		\\
54543	&	12.04 	&	0.08 	&	11.26 	&	0.07 	&	10.64 	&	0.07 	&	10.01 	&	0.05 	&		&		\\
54561	&	11.71 	&	0.07 	&	10.94 	&	0.07 	&	10.18 	&	0.06 	&	9.88 	&	0.05 	&	11.08 	&	0.07 	\\
54571	&	11.56 	&	0.05 	&	10.87 	&	0.07 	&	10.12 	&	0.06 	&	9.82 	&	0.05 	&	11.01 	&	0.07 	\\
54578	&	11.45 	&	0.05 	&	10.68 	&	0.08 	&	9.99 	&	0.07 	&	9.71 	&	0.06 	&	10.81 	&	0.08 	\\
54585	&	11.31 	&	0.06 	&	10.53 	&	0.08 	&	9.89 	&	0.07 	&	9.57 	&	0.04 	&	10.70 	&	0.07 	\\
54593	&	11.34 	&	0.05 	&	10.58 	&	0.07 	&	9.95 	&	0.06 	&	9.65 	&	0.05 	&	10.74 	&	0.07 	\\
54601	&	11.37 	&	0.05 	&	10.62 	&	0.08 	&	9.96 	&	0.06 	&	9.68 	&	0.04 	&	10.76 	&	0.07 	\\
54613	&	11.14 	&	0.05 	&	10.42 	&	0.08 	&	9.77 	&	0.07 	&	9.48 	&	0.05 	&	10.54 	&	0.07 	\\
54627	&	11.01 	&	0.05 	&	10.31 	&	0.07 	&	9.74 	&	0.07 	&	9.45 	&	0.05 	&	10.53 	&	0.07 	\\
54631	&	11.00 	&	0.05 	&	10.33 	&	0.07 	&	9.71 	&	0.06 	&	9.42 	&	0.05 	&	10.48 	&	0.07 	\\
54642	&	10.95 	&	0.05 	&	10.29 	&	0.07 	&	9.71 	&	0.08 	&	9.39 	&	0.05 	&	10.45 	&	0.08 	\\
54652	&	10.94 	&	0.05 	&	10.27 	&	0.07 	&	9.68 	&	0.06 	&	9.42 	&	0.08 	&	10.45 	&	0.07 	\\
54658	&	10.90 	&	0.05 	&	10.24 	&	0.07 	&	9.64 	&	0.06 	&	9.32 	&	0.04 	&	10.38 	&	0.07 	\\
54674	&	10.87 	&	0.05 	&	10.23 	&	0.07 	&	9.62 	&	0.07 	&	9.34 	&	0.05 	&	10.37 	&	0.07 	\\
54689	&	10.80 	&	0.05 	&	10.17 	&	0.10 	&	9.61 	&	0.07 	&	9.30 	&	0.08 	&	10.33 	&	0.07 	\\
54692	&	10.71 	&	0.05 	&	10.08 	&	0.07 	&	9.53 	&	0.07 	&	9.23 	&	0.05 	&	10.24 	&	0.07 	\\
54699	&	10.74 	&	0.06 	&	10.14 	&	0.07 	&	9.51 	&	0.06 	&		&		&	10.28 	&	0.08 	\\
54703	&	10.70 	&	0.07 	&	10.07 	&	0.09 	&	9.47 	&	0.07 	&	9.25 	&	0.05 	&	10.21 	&	0.08 	\\
54718	&	10.65 	&	0.07 	&	10.05 	&	0.09 	&	9.37 	&	0.06 	&	9.19 	&	0.04 	&	10.14 	&	0.07 	\\
54726	&	10.65 	&	0.05 	&	10.08 	&	0.08 	&	9.45 	&	0.13 	&	9.24 	&	0.05 	&	10.17 	&	0.07 	\\
54732	&	10.57 	&	0.05 	&	10.06 	&	0.09 	&	9.37 	&	0.08 	&	9.24 	&	0.04 	&	10.22 	&	0.09 	\\
54734	&	10.48 	&	0.39 	&	9.94 	&	0.28 	&	9.43 	&	0.06 	&	9.20 	&	0.11 	&	10.13 	&	0.13 	\\
54741	&	10.59 	&	0.05 	&	10.00 	&	0.07 	&	9.37 	&	0.08 	&	9.14 	&	0.07 	&	10.08 	&	0.07 	\\
54748 	&	10.60 	&	0.05 	&		&		&		&		&		&		&	10.15 	&	0.12 	\\
54872	&	10.48 	&	0.06 	&	9.86 	&	0.07 	&	9.34 	&	0.06 	&	9.04 	&	0.04 	&	10.01 	&	0.07 	\\
54881	&	10.55 	&	0.06 	&	10.01 	&	0.07 	&	9.37 	&	0.07 	&	9.09 	&	0.04 	&	10.10 	&	0.07 	\\
54903	&	10.66 	&	0.05 	&	10.07 	&	0.07 	&	9.45 	&	0.06 	&	9.15 	&	0.04 	&	10.15 	&	0.07 	\\
54929	&	10.86 	&	0.06 	&	10.22 	&	0.08 	&	9.46 	&	0.07 	&	9.28 	&	0.04 	&	10.37 	&	0.07 	\\
54946	&	10.91 	&	0.06 	&	10.28 	&	0.07 	&	9.56 	&	0.06 	&	9.26 	&	0.05 	&	10.37 	&	0.07 	\\
54953	&		&		&	10.27 	&	0.07 	&	9.61 	&	0.06 	&	9.29 	&	0.04 	&	10.39 	&	0.08 	\\
54961	&	10.95 	&	0.06 	&	10.29 	&	0.08 	&	9.66 	&	0.07 	&	9.36 	&	0.05 	&	10.44 	&	0.07 	\\
54978	&	10.95 	&	0.05 	&	10.30 	&	0.07 	&	9.61 	&	0.06 	&	9.33 	&	0.04 	&	10.45 	&	0.07 	\\
55069	&	11.60 	&	0.05 	&	10.92 	&	0.07 	&	10.16 	&	0.07 	&	9.84 	&	0.04 	&	11.08 	&	0.07 	\\
55268	&	11.22 	&	0.12 	&	10.60 	&	0.07 	&	9.94 	&	0.06 	&	9.54 	&	0.05 	&	10.73 	&	0.07 	\\
55293	&	11.26 	&	0.06 	&	10.57 	&	0.07 	&	9.84 	&	0.06 	&	9.54 	&	0.04 	&	10.73 	&	0.08 	\\
55317	&		&		&		&		&		&		&		&		&	10.69 	&	0.07 	\\
55319	&		&		&	10.48 	&	0.07 	&	9.74 	&	0.06 	&	9.44 	&	0.05 	&	10.62 	&	0.07 	\\
55322	&	11.13 	&	0.06 	&	10.47 	&	0.08 	&	9.75 	&	0.07 	&	9.44 	&	0.05 	&	10.63 	&	0.08 	\\
55358	&	11.12 	&	0.05 	&	10.49 	&	0.08 	&	9.80 	&	0.07 	&	9.48 	&	0.04 	&	10.63 	&	0.07 	\\
55399	&		&		&	10.36 	&	0.07 	&	9.66 	&	0.07 	&	9.36 	&	0.05 	&	10.47 	&	0.08 	\\
55402	&	11.07 	&	0.06 	&	10.45 	&	0.08 	&	9.73 	&	0.07 	&	9.48 	&	0.05 	&	10.57 	&	0.07 	\\
55411	&	10.90 	&	0.07 	&	10.32 	&	0.09 	&	9.65 	&	0.06 	&	9.34 	&	0.05 	&	10.45 	&	0.08 	\\
55414	&	11.03 	&	0.05 	&	10.37 	&	0.07 	&	9.66 	&	0.07 	&	9.35 	&	0.05 	&	10.50 	&	0.07 	\\
55426	&	11.11 	&	0.05 	&	10.46 	&	0.08 	&	9.73 	&	0.06 	&	9.46 	&	0.05 	&	10.60 	&	0.07 	\\
55449	&	11.16 	&	0.05 	&	10.53 	&	0.08 	&	9.81 	&	0.06 	&	9.50 	&	0.05 	&	10.69 	&	0.07 	\\
55657	&	11.05 	&	0.05 	&	10.46 	&	0.07 	&	9.80 	&	0.06 	&	9.60 	&	0.04 	&		&		\\
55663	&		&		&		&		&		&		&		&		&	10.65 	&	0.07 	\\
\hline
 \end{tabular} 
\end{table*}

\setcounter{table}{2}
\begin{table*} %Table 3 online
\caption[]{Journal of spectroscopic observations of V1280 Sco.}
\begin{tabular}{llccrcll}
\hline
\hline
Date & JD & Range & Resolution & Exp. & Observatory & Phase & Remarks\\
UT &  2400000+ & \AA\ & \AA & sec & & \\
\hline
2007 & & & & &\\
Feb. 5.87 & 54\,137.37 & 4100-6760 & 4.7 & 900 & NHAO & initial rising & Fig. 4$\dagger$\\
Feb. 12.88 & 54\,144.38 & 3750-8250 & 10 &  540 & FBO & pre-maximum halt & Fig. 5\\
Feb. 14.83 & 54\,146.33 & 3750-8250 & 10 &  900 & FBO & pre-maximum rising & \\
Feb. 14.86 & 54\,146.36 & 4160-6820 & 4.7 & 600 & NHAO & pre-maximum rising &Fig. 5\\
Feb. 15.86 & 54\,147.36 & 3750-8250 & 10 &  1380 & FBO & peak fluctuations & Fig. 5\\
Feb. 16.80 & 54\,148.30 & 4160-6820 & 4.7 & 600 & NHAO & peak fluctuations &$\dagger$\\
Feb. 18.85 & 54\,150.35 & 3750-8250 & 10 &  600 & FBO & peak fluctuations & Fig. 5\\
Feb. 19.84 & 54\,151.34 & 4160-6820 & 4.7 & 300 & NHAO & & Fig. 7$\ast$\\
Feb. 19.84 & 54\,151.34 & 3750-8250 & 10 &  1000 & FBO & &\\
Feb. 20.84 & 54\,152.34 & 3750-8250 & 10 &  470 & FBO & &\\
Feb. 21.82 & 54\,153.32 & 4160-6820 & 4.7 & 600 & NHAO & & Fig. 8$\ast$\\
Feb. 25.83 & 54\,157.33 & 4160-6820 & 4.7 & 300 & NHAO & &\\
Feb. 25.86 & 54\,157.36 & 3750-8250 & 10 &  144 & FBO & &\\
Feb. 27.84 & 54\,159.34 & 3750-8250 & 10 &  120 & FBO & dust phase &\\
Feb. 28.81 & 54\,160.31 & 4160-6820 & 4.7 & 300 & NHAO &  dust phase & Fig. 9\\
Mar. 6.85 & 54\,166.35 & 3750-8250 & 10 &  1440 & FBO & dust phase &\\
Mar. 7.80 & 54\,167.30 & 4160-6820 & 4.7 & 1800 & NHAO & dust phase &\\
Mar. 8.85 & 54\,168.35 & 3750-8250 & 10 &  1440 & FBO & dust phase &\\
Mar. 16.82 & 54\,176.32 & 3750-8250 & 10 &  1800 & FBO & dust phase &\\
Mar. 21.82 & 54\,181.32 & 4160-6820 & 4.7 & 1800 & NHAO & dust phase &\\
Mar. 25.80 & 54\,185.30 & 4160-6820 & 4.7 & 2400 & NHAO & dust phase & Fig. 10$\ast$\\
Apr. 11.82 & 54\,202.32 & 4160-6820 & 4.7 & 2400 & NHAO & dust phase &\\
May 2.78 & 54\,223.28 & 4160-6820 & 4.7 & 4800 & NHAO & dust phase &\\
May 7.74 & 54\,228.24 & 4160-6820 & 4.7 & 5400 & NHAO & dust phase &\\
May 20.69 & 54\,241.19 & 4160-6820 & 4.7 & 3600 & NHAO & dust phase (re-brightening)& Fig. 11\\
May 26.72 & 54\,247.22 & 4160-6820 & 4.7 & 3600 & NHAO & dust phase (re-brightening) &\\
Jun. 4.64 & 54\,256.14 & 4160-6820 & 4.7 & 3600 & NHAO & dust phase &\\
Jun. 10.63 & 54\,262.13 & 4160-6820 & 4.7 & 1800 & NHAO & dust phase &\\
\hline
2008 & & & & & &\\
Feb. 7.86 & 54\,504.36 & 3750-8250 & 10 &  2880 & FBO & plateau &\\
Feb. 18.87 & 54\,515.37 & 4160-6820 & 4.7 & 2400 & NHAO & plateau &\\
Feb. 27.78 & 54\,524.28 & 3750-8250 & 10 &  720 & FBO & plateau &\\
Jun. 9.67 & 54\,627.17 & 4160-6820 & 4.7 & 3600 & NHAO & plateau &\\
Jul. 8.58 & 54\,656.08 & 4160-6820 & 4.7 & 2400 & NHAO & plateau & \\
Jul. 30.57 & 54\,678.07 & 4160-6820 & 4.7 & 1800 & NHAO & plateau &\\
Sep. 8.44 & 54\,717.94 & 4160-6820 & 4.7 & 1800 & NHAO & plateau &$\dagger$\\
Sep. 9.43 & 54\,718.93 & 4160-6820 & 4.7 & 1800 & NHAO & plateau &Fig. 12$\ast$\\
\hline
2009 & & & & & &\\
Apr. 10.80  & 54\,932.30 & 3900-8200  & 5 & 1800 & BAO & plateau & Fig. 13\\
May 9.52	  & 54\,961.02 & 4130-6860  & 0.1 & 900 & Subaru & plateau & \\
Jun. 15.52 & 54\,998.02 & 4130-6860  & 0.1 & 600 & Subaru & plateau & \\
Jun. 16.50 & 54\,999.00 & 4130-6860 & 0.1 & 900 & Subaru & plateau & \\
Jun. 26.55 & 55\,009.05 & 4000-8000 & 10 & 900 & GAO & plateau &\\
Jul. 4.46     & 55\,016.96 &  4130-6860 & 0.1 & 900 & Subaru & plateau & \\
Jul. 6.40	 & 55\,018.90  & 4130-6860 & 0.1 & 1200 & Subaru & plateau & \\
Sep. 10.44 & 55\,084.94 & 4000-8000 & 10 & 480 & GAO & plateau &\\
Sep. 16.42 & 55\,090.92 & 4000-8000 & 10 & 1260 & GAO & plateau &\\
\hline
2010 & & & & & &\\
Mar. 11.80 & 55\,267.30 & 4120-9680  & 10 & 600 & HHAO & plateau & \\
Mar. 12.80 & 55\,268.30 & 4000-8000  & 10 & 1260 & GAO & plateau & Fig. 13\\
Apr. 17.70 & 55\,304.20 &  4130-9690 & 10 & 900 & HHAO & plateau & \\
Jun. 24.60 & 55\,372.10 & 4000-8000  & 10 & 600 & GAO & plateau & \\
Jul. 1.40     & 55\,378.90 & 4130-6860 & 0.1 & 600 &  Subaru & plateau &\\
Jul. 19.53 & 55\,397.03 & 3700-7500  & 10 & 1500 & KAO & plateau & Fig. 13\\
Aug. 1.51 & 55\,410.01 & 4100-9640  & 10 & 900 & HHAO & plateau & Fig. 13\\
\hline
2011  & & & & & &\\
Mar. 17.64 & 55\,638.14 & 4130-6860 & 0.1 & 600   &  Subaru & plateau & Fig. 14\\
Apr. 11.14 & 55\,662.64 & 4470-6890  & 5    & 600    & Asiago   & plateau & Fig. 13\\
Jun. 12.44  & 55\,724.94    & 4130-6860 &  0.1 &  600    &  Subaru & plateau & Fig. 14\\
Jul. 24.28  &  55\,766.78 & 4130-6860 &  0.1 &   600   &  Subaru & plateau & Fig. 14\\
\hline
\multicolumn{6}{l}{UT: Universal time at start of exposure}\\
\multicolumn{6}{l}{$\ast$: Spectra are shown in the online material}\\
\multicolumn{6}{l}{$\dagger$: Data without the flux calibration because of a poor weather condition or missing $y$-band observation}\\
 \end{tabular} 
\end{table*}

\setcounter{table}{4}
\begin{table*}[bht] %Table 5 online
\begin{flushbottom}
\scalebox{0.9}{
\rotatebox{90}{\begin{minipage}{\textheight}
\centering
\caption[]{Equivalent widths for absorption lines and fluxes for emission lines measured with the NHAO spectra.}
\begin{tabular}{lrrrrrrrrrrrrrrrrrrrrrr}
\hline
\hline
						&  \multicolumn{2}{c}{2007}	& \multicolumn{18}{l}{}		\\
Date	(UT)					& \multicolumn{2}{c}{Feb. 14.86}	& \multicolumn{2}{c}{Feb. 19.84}	& \multicolumn{2}{c}{Feb. 21.82}	& \multicolumn{2}{c}{Feb. 25.83}	& \multicolumn{2}{c}{Feb. 28.81}	& \multicolumn{2}{c}{Mar. 7.80}	& \multicolumn{2}{c}{Mar. 21.82}	& \multicolumn{2}{c}{Mar. 25.80}	& \multicolumn{2}{c}{Apr. 11.82}	& \multicolumn{2}{c}{May 2.78}\\
JD - 2\,400\,000				& \multicolumn{2}{c}{54\,146.36}	& \multicolumn{2}{c}{54\,151.34}	& \multicolumn{2}{c}{54\,153.32}	& \multicolumn{2}{c}{54\,157.33}	& \multicolumn{2}{c}{54\,160.31}	& \multicolumn{2}{c}{54\,167.30}	& \multicolumn{2}{c}{54\,181.32}	& \multicolumn{2}{c}{54\,185.30}	& \multicolumn{2}{c}{54\,202.32}	& \multicolumn{2}{c}{54\,223.28}\\				
\hline
ID						& abs & em	&abs & em	&abs & em	&abs & em	&abs & em	&abs & em	&abs & em	&abs & em	&abs & em	&abs & em\\ 
\hline
H$\delta$ 4102				&	&		&	&		& 0.90&	211.6&	&452.0	&	& 324.9	&	&		&	&		&	&		&	&		&	&	\\
\ion{Fe}{ii} (27,28) 4173/4178	& 1.21&		& 	&		&	&	101.6&	&258.8	&	&232.3	&	&2.62	&	&		&	& 0.27	&	&		&	&	\\
\ion{Fe}{ii} (27) 4233			& 0.54&		&	&		&	&	62.4	&	&161.3	&	&124.7	&	&2.42	&	&		&	& 0.54	&	&		&	&	\\
\ion{Fe}{ii} (27) 4273			& 0.28&		&	&		&	&	100.8&	&112.1	&	&146.6	&	&		&	&		&	&		&	&		&	&	\\
H$\gamma$ 4341			& 3.35&		& 	&		& 1.15&	186.6&	&523.7	&	&370.1	&	&3.52	&	&		&	&0.97	&	&		&	&	\\
\ion{Fe}{ii} (27) 4417			& 0.66&		&	&		&	&	74.5	&	&231.0	&	&129.0	&	&		&	&		&	&		&	&		&	&	\\
\ion{Fe}{ii} (38,37) 4508/4520	& 0.68&		&	&		&	&		&	&		&	&		&       &6.48	&	& 1.07	&	& 1.73	&	&		&	&	\\
\ion{Fe}{ii} (37) 4556			&	&		&	&		&	&		&	&		&	&		&	&3.92	&	& 1.29	&	& 1.33	&	&		&	&	\\
\ion{Fe}{ii} (38) 4584			&	&		&	&		&	&		&	&		&	&		&	&5.03	&	& 1.05	&	& 1.28	&	&		&	&	\\
\ion{Fe}{ii} (37) 4629			&	&		&	&		&	&		&	&		&	&		&	&2.15	&	& 0.57	&	& 0.56	&	&		&	&	\\
H$\beta$ 4861				& 2.61&		& 1.65&114.8	& 0.79&334.5	&	&988.0	&	&733.9	&	&7.85	&	&1.30	&	&1.57	&	&0.46	&	&	\\
\ion{Fe}{ii} (42) 4924			& 0.59&	1.61	& 2.64&91.1	& 1.28&142.9	&	&313.8	&	& 276.2	&	&4.06	&	&0.82	&	&1.07	&	&0.36	&	&	\\
\ion{Fe}{ii} (42) 5018			& 0.81&	1.48	& 3.07&79.0	& 3.51& 114.6	&	&366.8	&	&224.9	&	&3.78	&	& 0.95	&	& 1.16	&	&0.39	&	&	\\
\ion{Fe}{ii} (52) 5169			& 1.04& 2.15	& 3.66&34.1	& 2.85&88.2	&	&281.6	&	&175.7	&	&4.57	&	& 0.82	&	& 1.17	&	&0.47	&	&	\\
\ion{Fe}{ii} (49) 5198			&	&		&	&		&	&		&	&		&	&		&	&1.86	&	& 0.36	&	& 0.51	&	&0.09	&	&	\\
\ion{Fe}{ii} (49) 5234			& 0.60&		&	&		&	&132.4	&	& 297.6	&	& 186.0	&	&1.56	&	& 0.30	&	& 0.42	&	&0.08	&	&	\\
\ion{Fe}{ii} (49) 5276			& 0.39&		&	&		& 	&126.6	&	& 314.4	&	&  257.0	&	&1.74	& 	& 0.26	&	& 0.61	&	&0.24 	&	&	\\
\ion{Fe}{ii} (49) 5317			& 0.48&		&	&		&	&98.0	&	& 228.0	&	& 151.4	&	&2.54	&	& 0.47	& 	&0.58	&	&0.18	&	&	\\
\ion{Fe}{ii} (48) 5337			& 0.68&		&	&		&	&88.9	&	&137.7	&	& 35.9	&	&		&	&		&	&		&	&		&	&	\\
\ion{Fe}{ii} (48) 5363			&	&		&	&		&	&		&	&		&	& 73.0	&	&0.69	&	&		&	& 0.13	&	&		&	&	\\
\ion{Fe}{ii} (49) 5425			&	&		&	&		&	&		&	&		&	& 58.6	&	& 0.60	&	&		&	& 0.12	&	&		&	&	\\
\ion{Fe}{ii}	 (55)	5535			&	&		&1.66&11.2	&	&		&	&		&	&65.2	&	& 0.70	&	& 0.09	&	&0.15	&	& 0.04	&	&	\\
$$[\ion{O}{i}] 5577			&	&		&	&		&	&		&	&		& 	&42.5	&	& 1.41	&	& 0.29	&	& 0.45	&	& 0.13	&	&	\\
$$[\ion{N}{ii}] 5755			&	&		&	&		&	&		&	&		&	& 7.71	&	&		&	&		&	& 0.14	&	& 0.11	&	&	\\
\ion{He}{i} 5876			& 0.07&		&	&		&	&		&	&		&	&		&	&		&	&		&	&		&	&		&	&	\\
\ion{Na}{i} D2 5890, D1 5896 	& 2.05&		& 6.80&20.3	& 10.7&64.0	& 5.85& 140.2	& 6.00&87.8	&	& 2.19	&	& 0.52	&	&0.60	&	&0.29	&	& 0.18\\
\ion{Fe}{ii} (46) 5991			&	&		&	&		&	&		&	&		& 	&25.0	&	& 0.39	&	&		&	&		&	&		&	&	\\
\ion{Fe}{ii} (74) 6148	, 6149	&	&		& 	&		&	&		&	&		 &	& 6.92	&	& 0.61	&	& 0.08	&	& 0.14	&	&		&	&	\\
\ion{O}{i} 6156, 6157, 6158	&0.90&		&0.99&24.4	& 0.23& 101.9	& 	&131.8	&	&52.2	&	&		&	&		&	&		&	&		&	&	\\
\ion{Fe}{ii} (74) 6248			& 0.26&		&	&25.0	&	&68.0	&	&130.4	&	&68.4	& 	&1.21	&	&		& 	&0.21	&	&		&	&	\\
unid 6285					& 0.88&		& 0.90&		& 0.95&		& 1.45&		&	&		&	&		&	&		&	&		&	&		&	&	\\
$$[\ion{O}{i}] 6300			&	&		&	&		&	&		&	&		& 	&26.7	&	& 3.36	&	& 1.06	&	&1.39	&	& 0.88	&	& 0.57 \\
\ion{Si}{ii} 6347				& 0.81&		& 1.81&5.43	& 1.43&		& 0.89&		&	&		&	&		&	&		&	&		&	&		&	&	\\
$$[\ion{O}{i}] 6364			&	&		&	&		&	&		&	&		& 	&20.3	&	&1.26	&	& 0.37	&	& 0.49	&	& 0.34	&	& 0.22\\
\ion{Si}{ii} 6371				& 0.65&		& 0.89&11.3	&1.19&		& 0.75&		&	&		&	&		&	&		&	&		&	&		&	&	\\
\ion{O}{i} 6454, 6456 / \ion{Fe}{ii} (74) 6456 & 0.69&		& 0.97&15.2	&	& 58.6	&	& 66.9	&	& 70.7	&	&0.82	&	&0.13	&	& 0.18	&	&		&	&	\\
\ion{N}{i} 6482, 6483, 6484, 6485	&0.71&		&	&		&	&		&	&		&	&		&	&		&	&		&	&		&	&		&	&	\\
\ion{N}{i} 6521				&	&		&	&		&	&		&	&		&	&		&	&		&	&		&	&		&	&		&	&	\\
H$\alpha$ 6563			& 1.77&	4.45	& 3.70&217.9	&2.39&637.1	& 32.0& 2200	&	&3304	&	&50.8	&	& 10.6	&	& 14.3	&	&5.85	&	&1.64\\
\ion{N}{i} 6645				& 0.19&		&	&		&	&		&	&		&	&		&	&		&	&		&	&		&	&		&	&	\\
\ion{N}{i} 6653				&0.22 &		&	&		&	&		&	&		&	&		&	&		&	&		&	&		&	&		&	&	\\
\ion{He}{i} 6678			&0.03&		&	&		&	&		&	&		&	&		&	&		&	&		&	&		&	&		&	&	\\
\ion{N}{i} 6723				&0.22&		&	&		&	&		&	&		&	&		&	&		&	&		&	&		&	&		&	&	\\
\hline
\multicolumn{5}{l}{The unit of equivalent width is $\AA$.}\\
\multicolumn{5}{l}{The unit of flux is  $10^{-11}$\erg.}
\end{tabular}
\end{minipage}
}}
\end{flushbottom}
\end{table*}

\setcounter{table}{4}
\begin{table*}[bht] %Table 5 online
\begin{flushbottom}
\scalebox{0.9}{
\rotatebox{90}{\begin{minipage}{\textheight}
\centering
\caption[]{continued.}
\begin{tabular}{lrrrrrrrrrrrrrrrrrrrrrr}
\hline
\hline
						&  \multicolumn{2}{c}{2007}	& \multicolumn{8}{l}{}									& \multicolumn{2}{c}{2008} & \multicolumn{10}{l}{}\\
Date	(UT)					&  \multicolumn{2}{c}{May 7.74}	&  \multicolumn{2}{c}{May 20.69}	& \multicolumn{2}{c}{ May 26.72}	&  \multicolumn{2}{c}{Jun. 4.64}	&  \multicolumn{2}{c}{Jun. 10.63}	&  \multicolumn{2}{c}{Feb. 18.87}	&  \multicolumn{2}{c}{Jun. 9.67}	&  \multicolumn{2}{c}{Jul. 8.58}		&  \multicolumn{2}{c}{Jul. 30.57}	& \multicolumn{2}{c}{ Sep. 9.43}	& & \\
JD - 2\,400\,000				&  \multicolumn{2}{c}{54\,228.24}	& \multicolumn{2}{c}{54\,241.19}	& \multicolumn{2}{c}{54\,247.22}	& \multicolumn{2}{c}{54\,256.14}	& \multicolumn{2}{c}{54\,262.13}	& \multicolumn{2}{c}{54\,515.37}	& \multicolumn{2}{c}{54\,627.17}	& \multicolumn{2}{c}{54\,656.08}	& \multicolumn{2}{c}{54\,678.07}	& \multicolumn{2}{c}{54\,718.93}	& & \\							
\hline
ID						& abs & em	& abs & em	&abs & em	&abs & em	&abs & em	&abs & em	&abs & em	&abs & em	&abs & em	&abs & em	& &\\ 
\hline
H$\delta$ 4102				&	&		&	&		&	&		& 	&		&	&		&	&		&	&		& 	&1.52	&	&		&	&1.25	& &\\
\ion{Fe}{ii} (27,28) 4173/4178	&	&		&	&		& 	&		&	&		&	&		&	&		&	&		&	&		&	&		&	&		& &\\
\ion{Fe}{ii} (27) 4233			&	&		&	&		& 	&		&	&		&	&		&	&		&	&		&	&		&	&		&	&		& &\\
\ion{Fe}{ii} (27) 4273			&	&		&	&		& 	&		&	&		&	&		&	&		&	&		&	&		&	&		&	&		& &\\
H$\gamma$ 4341			&	&		&	&		& 	&		& 	&		&	&		&	&		&	&		&1.54&1.98	&	& 2.42	&2.56& 3.12	& &\\
\ion{Fe}{ii} (27) 4417			&	&		&	&		& 	&		&	&		&	&		&	&		&	&		&	&		&	&		&	&		& &\\
\ion{Fe}{ii} (38,37) 4508/4520 	&	&		&	&		& 	&		&	&		&	&		&	&		&	&		&	&		&	&		&	&		& &\\
\ion{Fe}{ii} (37) 4556			&	&		&	&		& 	&		&	&		&	&		&	&		&	&		&	&		&	&		&	&		& &\\
\ion{Fe}{ii} (38) 4584			&	&		&	&		& 	&		&	&		&	&		&	&		&	&		&	&		&	&		&	&		& &\\
\ion{Fe}{ii} (37) 4629			&	&		&	&		& 	&		&	&		&	&		&	&		&	&		&	&		&	&		&	&		& &\\
H$\beta$ 4861				&	&		&	&0.54	& 6.89&1.13	& 	&		&	&		&	&3.06	&	&3.19	& 0.87&4.24	& 1.41&4.82	&1.70&4.49	& &\\
\ion{Fe}{ii} (42) 4924			& 	&		&	& 0.22	& 	&0.46	& 	&0.13	&	&		&	&		&	&		& 0.30&0.21	& 0.44&0.23	&0.27&0.38		& &\\
\ion{Fe}{ii} (42) 5018			& 	&		& 	&0.42	& 	&0.54	& 	&0.18	&	&		&	&		&	&		& 0.19& 0.23	&	&0.25	&0.15&0.34		& &\\
\ion{Fe}{ii} (52) 5169			&	&		& 	&0.54	& 	&0.66	&	&0.23	&	&		&	&		&	&		&0.32& 0.17	& 0.30& 0.14	&0.29&0.19		& &\\
\ion{Fe}{ii} (49) 5198			&	&		&	&		& 	&		&	&		&	&		&	&		&	&		&	&		&	&		&	&		& &\\
\ion{Fe}{ii} (49) 5234			&	&		&	&		& 	&		&	&		&	&		&	&		&	&		&	&0.04	&	& 0.04	&	&0.08		& &\\
\ion{Fe}{ii} (49) 5276			&	&		&	&		& 	&		&	&		&	&		&	&		&	&		&	& 0.10	&	& 0.10	&	&0.16		& &\\
\ion{Fe}{ii} (49) 5317			& 	&		& 	& 0.18	&	&0.21	&	&		&	&		&	&		&	&		& 0.13&0.08	& 	&0.09	&0.06&0.16		& &\\
\ion{Fe}{ii} (48) 5337			&	&		&	&		& 	&		&	&		&	&		&	&		&	&		&	&		&	&		&	&		& &\\
\ion{Fe}{ii} (48) 5363			&	&		&	&		& 	&		&	&		&	&		&	&		&	&		&	&		&	&		&	&0.06		& &\\
\ion{Fe}{ii} (49) 5425			&	&		&	&		& 	&		&	&		&	&		&	&		&	&		&	&		&	&		&	&		& &\\
\ion{Fe}{ii}	 (55)	5535			&	&		&	&		& 	&		&	&		&	&		&	&		&	&		&	&0.06	&	&0.05	&	&0.11	& &\\
$$[\ion{O}{i}] 5577			&	&		&	&		& 	&		&	&		&	&		&	&		&	&		&	&		&	&		&	&		& &\\
$$[\ion{N}{ii}] 5755			&	&		&	&		&	&		&	&		&	&		& 	&		&	&		&	&0.81	&	& 0.74	&	&0.95	& &\\
\ion{He}{i} 5876			&	&		&	&		& 	&		&	&		&	&		&	&		&	&		&	&0.22	&	& 0.24	&	&0.27	& &\\
\ion{Na}{i} D2 5890, D1 5896	& 	&0.11	& 11.6&0.20	&  9.92&0.32	& 	&		& 	&		&	&		&	&		& 	&		&	&		&	&		& &\\
\ion{Fe}{ii} (46) 5991			&	&		&	&		& 	&		&	&		&	&		&	&		&	&		&	&		&	&		&	&		& &\\
\ion{Fe}{ii} (74) 6148	, 6149		&	&		&	&		& 	&		&	&		&	&		&	&		&	&		&	&		&	&		&	&		& &\\
\ion{O}{i} 6156, 6157, 6158	&	&		&	&0.09	& 	&		&	&		&	&		&	&		&	&		&	&		&	&		&	&		& &\\
\ion{Fe}{ii} (74) 6248			&	&		&	&0.04	& 	&		&	&		&	&		&	&		&	&		&	&		&	&		&	&		& &\\
unid 6285					&	&		&	&		& 	&		&	&		&	&		&	&		&	&		&	&		&	&		&	&		& &\\
$$[\ion{O}{i}] 6300			& 	&0.36	&	& 0.32	&	& 0.46	&	& 0.48	&	& 0.18	&	&		&	& 		&	& 		&	& 		&	& 		 & &\\
\ion{Si}{ii} 6347				&	&		&	&		& 	&		&	&		&	&		&	&		&	&		&	&		&	&		&	&		& &\\
$$[\ion{O}{i}] 6364			&	& 0.13	&	& 0.12	&	& 0.20	&	& 0.15	&	& 0.08	&	& 		& 	& 		&	& 		&	&		&	& 		& &\\
\ion{Si}{ii} 6371				&	&		&	&		& 	&		&	&		&	&		&	&		&	&		&	&		&	&		&	&		& &\\
\ion{O}{i} 6454, 6456 / \ion{Fe}{ii} (74) 6456&	&		&	&		& 	&		&	&	&	&		&	&		&	&		&	&		&	&		&	&		& &\\
\ion{N}{i} 6482, 6483, 6484, 6485	&	&		&	&		& 	&		&	&		&	&		&	&		&	&		&	&		&	&		&	&		& &\\
\ion{N}{i} 6521				&	&		&	&		& 	&		&	&		&	&		&	&		&	&		&	&		&	&		&	&		& &\\
H$\alpha$ 6563			& 	&2.71	&  2.85&3.18	& 6.85&5.36	&	&2.64	& 	&0.90	&	&13.4	&	&16.2	&	&16.7	&	&21.0	&	&22.9	& &\\
\ion{N}{i}  6645				&	&		&	&		& 	&		&	&		&	&		&	&		&	&		&	&		&	&		&	&		& &\\
\ion{N}{i}	6653				&	&		&	&		& 	&		&	&		&	&		&	&		&	&		&	&		&	&		&	&		& &\\
\ion{He}{i} 6678			&	&		&	&		& 	&		&	&		&	&		&	&		&	&		&	&0.16	&	&0.20	&	&0.22	& &\\
\ion{N}{i} 6723				&	&		&	&		& 	&		&	&		&	&		&	&		&	&		&	&		&	&		&	&		& &\\
\hline
\multicolumn{5}{l}{The unit of equivalent width is $\AA$.}\\
\multicolumn{5}{l}{The unit of flux is  $10^{-11}$\erg.}
\end{tabular}
\end{minipage}
}}
\end{flushbottom}
\end{table*}

\end{document}